\documentclass[11pt,onecolumn,amssymb,nofootinbib,showpacs]{revtex4}
\usepackage{amsmath, amsthm, amscd, amssymb}
\usepackage{bm}
\usepackage{bbm}
\usepackage{graphicx}

\begin{document}

\title{\bf Non-Markovian dynamics for a free quantum particle
subject to spontaneous collapse in space: general solution and main
properties}
\author{Angelo Bassi}
\email{bassi@ts.infn.it}
\affiliation{Dipartimento di Fisica Teorica,
Universit\`a di Trieste, Strada Costiera 11, 34014 Trieste, Italy.
 \\ Istituto Nazionale di Fisica Nucleare,
Sezione di Trieste, Strada Costiera 11, 34014 Trieste, Italy.}
\author{Luca Ferialdi}
\email{ferialdi@ts.infn.it}
\affiliation{Dipartimento di Fisica Teorica,
Universit\`a di Trieste, Strada Costiera 11, 34014 Trieste, Italy.
\\ Istituto Nazionale di Fisica Nucleare, Sezione di Trieste, Strada
Costiera 11, 34014 Trieste, Italy.}
\begin{abstract}
We analyze the non-Markovian dynamics of a quantum system subject to
spontaneous collapse in space. After having proved, under suitable
conditions, the separation of the center-of-mass and relative
motions, we focus our analysis on the time evolution of the center
of mass of an isolated system (free particle case). We compute the
explicit expression of the Green's function, for a generic Gaussian
noise, and analyze in detail the case of an exponential correlation
function. We study the time evolution of average quantities, such as
the mean position, momentum and energy. We eventually specialize to
the case of Gaussian wave functions, and prove that all basic facts
about collapse models, which are known to be true in the white noise
case, hold also in the more general case of non-Markovian dynamics.
\end{abstract}
 \maketitle

\section{Introduction}
\label{sec:one}

Among the several models of spontaneous wave function collapse
which has been considered so far, the so--called QMUPL (Quantum
Mechanics with Universal Position Localization) model is
particularly interesting, being it a very good compromise between
mathematical simplicity and physical adequacy. It was first
introduced by Diosi~\cite{Diosi:89,Diosi:90} and subsequently
studied
in~\cite{Belavkin:89,Belavkin:92,Chruscinski:92,Gatarek:91,Halliwell:95,Holevo:96,Bassi2:05,Bassi:08,Bassi5:08},
both from the mathematical as well as physical point of view. It is
particularly relevant because it is the simplest model describing
the evolution of the wave function of a system of $N$
distinguishable particles, subject to a spontaneous collapse in
space; as such, it can be analyzed in great mathematical detail. The
model is defined by the following stochastic differential equation
(SDE):
\begin{equation} \label{eq:nlemp}
d\,\psi_{t}(\{ x \}) =  \left[ -\frac{i}{\hbar}\, H\, dt +
\sum_{n=1}^{N}\sqrt{\lambda_{n}}\, ( q _{n} - \langle q _{n}
\rangle_{t})\, d W_{n,t} - \frac{1}{2}\,\sum_{n=1}^{N} \lambda_{n} (
q _{n} - \langle  q _{n} \rangle_{t})^2 dt \right] \psi_{t}(\{ x
\});
\end{equation}
where the symbol $\{ x \} \equiv x_{1}, x_{2}, \ldots x_{n}$ denotes
the coordinates of the $N$ particles (for simplicity, we will work
in one spatial dimension). The operator $H$ is the standard quantum
hamiltonian of the composite system; $q_{n}$ is the position
operator associated to the $n$-th particle and $\langle q _{n}
\rangle_{t} \equiv \langle \psi_{t} | q_{n} | \psi_{t} \rangle$
denotes the quantum expectation value of $q_{n}$. The stochastic
processes $W_{n,t}$ are $N$ independent standard Wiener processes
defined on a probability space $(\Omega, {\mathcal F}, {\mathbb
P})$, and the parameters $\lambda_{n}$ are $N$ positive coupling
constants which is convenient to take proportional to the mass of
the particle according to the formula~\cite{Bassi2:05}:
\begin{equation}
\lambda_{n} \; = \; \frac{m_{n}}{m_{0}}\, \lambda_{0},
\end{equation}
where $m_{n}$ is the mass of the $n$-th particle, $m_{0}$ is a
reference mass which, at the non relativistic level, is reasonable
to take equal to the mass of a nucleon ($m_{0} \simeq 1.67 \times
10^{-27}$ Kg), while $\lambda_{0}$ is the only true new parameter of
the model, whose value sets the strength of the collapse mechanism.

The numerical value of $\lambda_{0}$ has to be chosen in such a way
that: i) the model reproduces quantum mechanical predictions for
microscopic systems; ii) it rapidly induces the collapse of the wave
function describing the center of mass of a macroscopic object. In
the literature, two quite different values for $\lambda_{0}$ have
been proposed, the first by Ghirardi, Rimini, and
Weber~\cite{Ghirardi:86}, and the second by Adler~\cite{Adler3:07}:
\begin{eqnarray}
\lambda_{0}^{\text{\tiny GRW}} & \simeq & 5.00\times 10^{-3} \;\;
\text{m$^{-2}$
sec$^{-1}$}, \label{eq:cvxvxc} \\
\lambda_{0}^{\text{\tiny Adler}} & \simeq & 1.12\times10^{6 \pm 2} \;
\text{m$^{-2}$ sec$^{-1}$}. \label{eq:cvxvxc2}
\end{eqnarray}
GRW's choice is motivated by the requirement that superpositions of
order $6.02\times 10^{23}$ nucleons, displaced by a distance of at
least $r_C = 10^{-5}$ cm, be suppressed within $10^{-3}$ sec.
Adler's choice instead is motivated by the requirement that the
collapse occurs already at the level of latent image formation. More
specifically, GRW set $\lambda_{\text{\tiny GRW}} \simeq 10^{-16}$
sec$^{-1}$, where $\lambda_{\text{\tiny GRW}}$ is the collapse rate
of the GRW model~\cite{Ghirardi:86}; Adler instead set
$\gamma_{\text{\tiny CSL}} \simeq 2\times 10^{-21\pm 2}$ cm$^{3}$
sec$^{-1}$~\cite{Adler3:07}, where $\gamma_{\text{\tiny CSL}}$ is
the noise-strength coupling constant of the CSL model; the relation
between $\lambda_0$ of our model and $\lambda_{\text{\tiny GRW}}$
and $\gamma_{\text{\tiny CSL}}$ is~\cite{Ghirardi2:90,Bassi2:05}:
$\lambda_0 = \alpha\lambda_{\text{\tiny GRW}}/2 =
\alpha^{5/2}\gamma_{\text{\tiny CSL}}/16 \pi^{3/2}$, with $\alpha =
1/r_C^2$.

Eq.~\eqref{eq:nlemp} is manifestly not linear, which makes it
difficult to analyze, in particular it makes it hard to find its
solutions. The way to circumvent this obstacle is to linearize the
equation, according to the following prescription~\cite{Holevo:96}. Consider
the linear SDE:
\begin{equation} \label{eq:lemp}
d\,\phi_{t}(\{ x \}) =  \left[ -\frac{i}{\hbar}\, H\, dt +
\sum_{n=1}^{N}\sqrt{\lambda_{n}}\, q _{n} \, d \xi_{n,t} -
\frac{1}{2}\,\sum_{n=1}^{N} \lambda_{n} q _{n}^2 dt \right]
\phi_{t}(\{ x \}),
\end{equation}
where the stochastic processes $\xi_{n,t}$ are $N$ independent
standard Wiener processes with respect to a new measure ${\mathbb
Q}$. It can be shown that $\| \psi_{t} \|^2$ is a
martingale, which can be used to generate a new measure from a given
one: the measure ${\mathbb Q}$ introduced here above is chosen in
such a way that $\| \psi_{t} \|^2$ is the Radon-Nikodyn derivative
of ${\mathbb P}$ with respect to ${\mathbb Q}$, i.e.: $d{\mathbb P}
= \| \phi_{t} \|^2 d{\mathbb Q}$. Moreover, Girsanov's
theorem~\cite{Lipster:01} states that the two sets of Wiener
processes $\{ W_{n,t} \}$ and $\{ \xi_{n,t} \}$ are related as
follows: $d W_{n,t} = d \xi_{n,t} - 2 \sqrt{\lambda} \langle q_{n}
\rangle_{t} dt$.

Given these ingredients, it is easy to relate the solutions of
Eq.~\eqref{eq:lemp} to those of Eq.~\eqref{eq:nlemp}: given a
solution $\phi_{t}$ of Eq.~\eqref{eq:lemp}, one first considers the
normalized state $\psi_{t} = \phi_{t}/ \| \phi_{t}\|$, and then
replaces the noises $\{ \xi_{n,t} \}$ with $\{ W_{n,t} \}$ according
to the formula given above. It is not difficult to show that the
wave function so obtained solves Eq.~\eqref{eq:nlemp}. Further
details can be found in~\cite{Holevo:96}.

The aim of this paper is to analyze the generalization of the QMUPL
model to non-Markovian Gaussian random processes. As discussed
in~\cite{Diosi:97,Diosi:98,Bassi:02}, the generalization of the
linear Eq.~\eqref{eq:lemp} to the non-Markovian case takes the form:
\begin{equation} \label{eq:main}
\frac{d}{dt}\, \phi_{t}(\{ x \}) = \left[ -\frac{i}{\hbar} H +
\sum_{n=1}^{N} \sqrt{\lambda_{n}} q_{n} w_{n}(t) - 2
\sum_{n,m=1}^{N} \sqrt{\lambda_{n}} q_{n} \int_{0}^{t} ds
D_{nm}(t,s) \frac{\delta}{\delta w_{m}(s)} \right] \phi_{t}(\{ x
\}),
\end{equation}
where the noises $w_{n}(t)$ are now supposed to be {\it Gaussian
random processes} whose mean and correlation function, expressed
with respect to the measure ${\mathbb Q}$, are equal to:
\begin{equation}
{\mathbb E}_{\mathbb Q} [ w_{n}(t) ] = 0, \qquad\quad {\mathbb
E}_{\mathbb Q} [ w_{n}(t) w_{m}(s) ] = D_{nm}(t,s).
\end{equation}
The correlation function is assumed to be real, symmetric and
positive-semidefinite~\cite{Gikhman:96}.

Like in the white-noise case, Eq.~\eqref{eq:main} does not preserve
the norm of the wave function. Accordingly, we assume that the {\it
physical states} are the normalized states:
\begin{equation}
\psi_{t}(\{ x \}) \; \equiv \; \frac{\phi_{t}(\{ x \})}{\| \phi_{t}
\|};
\end{equation}
moreover, we assume that the {\it physical probability} measure is
${\mathbb P}$, which is related to the measure ${\mathbb Q}$, by
means of which the statistical properties of the noises $w_n(t)$
have been defined, according to the formula:
\begin{equation} \label{eq:vvbp}
d {\mathbb P} \; \equiv \; \| \phi_{t} \|^2 d {\mathbb Q}.
\end{equation}
{\it Measurable quantities} are given by expressions of the form: ${\mathbb E}_{\mathbb P}[
\langle \psi_t | O | \psi_t \rangle ]$, where $O$ is a suitable self
adjoint operator. According to relation~\eqref{eq:vvbp}, the
following equality holds true:
\begin{equation}
{\mathbb E}_{\mathbb P}[ \langle \psi_t | O | \psi_t \rangle ] \;
\equiv \; {\mathbb E}_{\mathbb P}[ \langle \phi_{t} | O | \phi_t
\rangle ].
\end{equation}
We then have the following useful results: physical properties can
be computed directly from the solution of the linear
equation~\eqref{eq:main}, if one averages with respect to the
measure ${\mathbb Q}$, instead of the physical measure ${\mathbb
P}$. For this reason, in the following we will focus our attention
only on Eq.~\eqref{eq:main}, without trying to investigate the form
of the corresponding non-linear but norm-preserving equation. This
will be the subject of future analysis.

The goal of the paper is to unfold the dynamics described by
Eq.~\eqref{eq:main}. We will focus our attention on the free
particle case, which can be solved exactly. The results can be
generalized to the case of harmonic oscillators, and in general to
any equation of the type~\eqref{eq:main}, containing expressions
which are at most quadratic in the operators $q$ and $p$. For other
type of systems, perturbation techniques can be
employed~\cite{Yu:99,Adler:07,Adler:08}. We list
the main results we have obtained. \vskip 0.2cm

\noindent \textsc{Result 1} (Sec.~\ref{sec:two}). Under the
conditions:
\begin{equation}
\sum_{m=1}^{N} \left[ \sqrt{\lambda_{m}} D_{mn}(t,s) -
\sqrt{\frac{\lambda_{n}\lambda_{m}}{\lambda_{N}}}\, D_{mN}(t,s)
\right] \; = \; 0, \qquad n=1,2, \ldots N-1,
\end{equation}
the center-of-mass motion and the relative motion decouple. In
particular, when $D_{nm}(t,s) = \delta_{nm}D(t,s)$, then the two
equations for the center-of-mass wave function
$\phi_{t}^{\text{\tiny cm}}$ and relative-motion wave function
$\phi_{t}^{\text{\tiny rel}}$ are:
\begin{equation}\label{eq:cm0}
\frac{d}{dt}\, \phi_{t}^{\text{\tiny cm}} = \left[ -\frac{i}{\hbar}
H_{\text{\tiny cm}} + \sqrt{\lambda}\, Q\, w(t) - 2 \sqrt{\lambda} Q
\int_{0}^{t} ds D(t,s) \frac{\delta}{\delta w(s)} \right]
\phi_{t}^{\text{\tiny cm}},
\end{equation}
\begin{equation} \label{eq:klfgf0}
\frac{d}{dt}\, \phi_{t}^{\text{\tiny rel}} = \left[ -\frac{i}{\hbar}
H_{\text{\tiny rel}} + \sum_{n=1}^{N} \sqrt{\bar{\lambda}_{n}}
\bar{q}_{n} \bar{w}_{n}(t) - 2 \sum_{n,m=1}^{N-1}
\sqrt{\bar{\lambda}_{n}} \bar{q}_{n}C_{nm} \int_{0}^{t} ds D(t,s)
\frac{\delta}{\delta \bar{w}_{m}(s)} \right] \phi_{t}^{\text{\tiny
rel}},
\end{equation}
where $w(t)$ and $\lambda$ are defined in~\eqref{eq:wnoise}
and~\eqref{eq:ftghdf}, and $\bar{w}_n(t)$, $\bar{\lambda}_n$ are
defined in~\eqref{eq:wbar} and~\eqref{eq:lbar}.\vskip 0.2cm

\noindent \textsc{Result 2} (Sec.~\ref{sec:three}). The Green's
function associated to a {\it free particle} of mass $m$ ($H=p^2/2m$) is:
\begin{equation}\label{eq:prop0}
G(x,t;x_0,0)= \sqrt{\frac{m}{2i\pi\hbar\, t\, u(t)}}
\exp\left[-\frac{im}{2\hbar}\left(x_0 z'(0) - x z'(t)\right)
+\frac{\sqrt{\lambda}}{2}\int_0^t ds\, w(s)z(s)\right],
\end{equation}
where $z(t)$ solves Eq.~\eqref{eq:z}, with boundary conditions:
$z(0) = x_0$, $z(t) = x$, while $u(t)$ is defined in
Eq.~\eqref{eq:det}. In particular, under the condition $D(t,s) =
D(|t-s|)$, which represents time translation invariance of the noise
$w(t)$, the Green's function~\eqref{eq:prop0} has the simpler
structure:
\begin{equation}
G(x,t;x_0,0)=\sqrt{\frac{m}{2i\pi\hbar\, t\,
u(t)}}\exp\left[-\mathcal{A}_t
(x_0^2+x^2)+\mathcal{B}_tx_0x+\mathcal{C}_tx_0+\mathcal{D}_tx+\mathcal{E}_t\right]\,,
\end{equation}
where:
\begin{equation}
\mathcal{A}_t \; = \; kf_t'(0), \quad\quad \mathcal{B}_t \; = \;
2kf_t'(t), \qquad\qquad k := \frac{i\hbar}{2m}\,,
\end{equation}
are deterministic coefficients, while:
\begin{eqnarray}
\mathcal{C}_t & = & -k h_t'(0) + \frac{\sqrt{\lambda}}{2}\int_0^t dl\,
w(l)f_t(l),\\
\mathcal{D}_t & = &  k h_t'(t) + \frac{\sqrt{\lambda}}{2}\int_0^t dl\, w(l) f_t(t-l), \\
\mathcal{E}_t & = & \frac{\sqrt{\lambda}}{2}\int_0^t dl\, w(l) h_t(l)
\,,
\end{eqnarray}
are random coefficients. The function $f(s)$ is solution of
Eq.~\eqref{eq:fans}, with boundary conditions $f(0) = 1$, $f(t) =
0$, while $h(s)$ is defined in Eq.~\eqref{eq:gfhsd}.  As we see, in
the non-Markovian time translation invariant case the free-particle
propagator has the same structure as in the white noise case. \vskip
0.2cm

\noindent \textsc{Result 3} (Sec.~\ref{sec:four}). Using the Green's
function~\eqref{eq:prop0}, one can rewrite the non-Markovian SDE for
a free quantum particle in a simpler way, replacing the functional
derivative with combinations of the position and momentum operators.
The collapsing equation takes the form:
\begin{equation}
\frac{d}{dt}\phi_t(x)=\left[ -\frac{i}{\hbar}
\frac{p^2}{2m} + \sqrt{\lambda} q w(t)- 2\lambda q \int_0^t ds
D(t,s)\big(q\,a_t(s)+p\,b_t(s)+c_t(s)\big)\right]\phi_t(x)\,,
\end{equation}
where $a_t(s)$, $b_t(s)$, $c_t(s)$ are defined
in~\eqref{eq:xvhkg0}--\eqref{eq:xvhkg2}. This result agrees with the Ansatz
first introduced in Sec. 4.3 of Ref.~\cite{Diosi:98}. \vskip 0.2cm

\noindent \textsc{Result 4} (Secs.~\ref{sec:five}
and~\ref{sec:six}). In the {\it white noise} limit, one recovers
well-known results in literature. More interestingly, we have solved explicitly the case of an {\it exponential correlation
function}:
\begin{equation}
D(t,s) \; = \; \frac{\gamma}{2}e^{-\gamma |t-s|},
\end{equation}
in which case the functions $f(s)$ and $h(s)$, through which
the coefficients ${\mathcal A}_t$--${\mathcal E}_t$ of the Green's
function can be computed, are:
\begin{equation}
f(s) = \frac{\sum_k\left[r_t^k \sinh \upsilon_k(t-s) +
u_t^k \cosh \upsilon_k(t-s) - u_s^k\right]}{ \sum_{k}\left[2c + r_t^k \, \sinh \upsilon_k t
+ u_t^k \cosh \upsilon_k t\right]}\,,
\end{equation}
while $h(s)$ is given by~\eqref{eq:gfhsd} with
\begin{eqnarray}
h^{\text{\tiny P}}(s)& =& -\frac{i\sqrt{\lambda}\hbar}{m}\int_0^s
\bar{f}_s(l)\left(w''(l)-\gamma^2w(l)\right)dl\,,\\
\bar{f}_s(l)&=&-\frac{1}{\zeta}\sum_k(-1)^k\frac{\sinh
\upsilon_k(s-l)}{\upsilon_k}\,.
\end{eqnarray}
\vskip 0.2cm

\noindent \textsc{Result 5} (Sec.~\ref{sec:seven}). We have proven
the so called {\it imaginary noise trick}, which we use  to compute
the average value of physical quantities, in particular the average
position $q$, average momentum $p$, and average kinetic energy $H_0
= p^2/2m$:
\begin{eqnarray}
\frac{d}{dt}{\mathbb E}_{\mathbb Q} \left[\langle q\rangle_t\right]
& = & \frac{1}{m}{\mathbb E}_{\mathbb Q} \left[\langle
p\rangle_t\right]\,, \\
\frac{d}{dt}{\mathbb E}_{\mathbb Q} \left[\langle p\rangle_t\right]
& = & 0\,,\\
\frac{d}{dt}{\mathbb E}_{\mathbb Q} \left[\langle H_0
\rangle_t\right] &=&\frac{\lambda\hbar^2}{m}\int_0^t ds D(t,s)\,.
\end{eqnarray}
The above equation prove that a free quantum particle moves, in the
average, like a classical particle, but its kinetic energy is not
conserved, not even in the average. This is an expected feature of
the model~\cite{Adler:07}.

\vskip 0.2cm

\noindent \textsc{Result 6} (Sec.~\ref{sec:eight}). We have studied
the time evolution of Gaussian wave functions,  analyzing in
particular the time evolution of the spread in position, and how the
fluctuations around the average of both the mean position and mean
momentum scale with the mass of the particle. In all cases, we have
recovered the results which hold true for the white noise models.

Before concluding this introductory section, two comments are
at order. First, it may seem rather remarkable that we have been
able to compute the Green's function associated to a non-Markovian
equation. In fact, dynamics with memory terms in general do now
allow for solutions expressed in terms of a Green's function, whose
nature is strictly linked to the Markovian character of the
evolution. The fact that in our case we have been able to compute
the Green's function appears less surprising if one looks at the way
in which Eq.~\eqref{eq:main} has been derived: originally, the
non-Markovian evolution was presented only in terms of a propagator
(see the pioneering works of Feynman and Vernon on decoherence~\cite{Feynman:63}
and of P. Pearle~\cite{Pearle:96,Pearle:96bis} on collapse models) and only afterwards a
non-Markovian equation was devised, having that propagator as the
associated Green's function~\cite{Strunz:96,Diosi:97}.

Second, it may appear odd that the integration interval in
Eq.~\eqref{eq:main} starts from 0 and not from $-\infty$, as
required by a truly non-Markovian dynamics. Here we are implicitly
making the assumption that the state at time $s=0$ suffices to
compute the subsequent evolution, without any need to know also the
previous history of the system. This assumption, which actually is
an approximation if the equation is taken as a fundamental equation,
is based on the following argument. We assume that before $s=0$ the
system had enough time to reach some equilibrium state, which does
not depend on the way it has been reached; this is the physical
argument behind the idea that the history prior to $s=0$ is
unimportant. Then, at time $s=0$ a sudden change in the system
occurs---e.g., due to a measurement-like interaction:
Eq.~\eqref{eq:main} describes how the system evolves thereafter.

\section{Separation of the center-of-mass and relative degrees of
freedom} \label{sec:two} Let us suppose that the total Hamiltonian
$H$ of Eq.~\eqref{eq:main}  separates in a term referring only to
the center of mass and one referring to the relative degrees of
freedom: $H \; = \; H_{\text{\tiny cm}} + H_{\text{\tiny rel}}$.
In this section we analyze under which conditions the dynamics of
the two types of motion decouple from each other. To this end, let
us  introduce the center of mass position operator $Q$ and the
relative position operators $\tilde{q}_{n}$ defined as follows:
\begin{eqnarray}
Q & \equiv & \displaystyle \frac{1}{M} \sum_{n=1}^{N} m_{n} \,
q_{n},
\qquad M \; \equiv \; \sum_{n=1}^{N} m_{n}\,, \\
& & \nonumber \\
\tilde{q}_{n} & \equiv & q_{n} - Q.
\end{eqnarray}
We will choose $Q$ and $\tilde{q}_{1}, \tilde{q}_{2}, \ldots,
\tilde{q}_{N-1}$ as a set of independent variables, while
$\tilde{q}_{N}$ is given in terms of the other operators according
to the formula:
\begin{equation}
\tilde{q}_{N} \; = \; - \frac{1}{m_{N}}\, \sum_{n=1}^{N-1} m_{n}
\tilde{q}_{n}.
\end{equation}

The second term on the right hand side of Eq.~\eqref{eq:main} trivially
separates in a term depending only on $Q$ and a term depending
only on the relative variables, as the following equality holds
true:
\begin{equation} \label{eq:dfgf}
\sum_{n=1}^{N} \sqrt{\lambda_{n}}\, q_{n}\, w_{n}(t) \; = \; Q
\sum_{n=1}^{N} \sqrt{\lambda_{n}}  w_{n}(t) + \sum_{n=1}^{N-1}
\sqrt{\lambda_{n}} \tilde{q}_{n}\left( w_{n}(t) -
\frac{\lambda_{n}}{\lambda_{N}}\, w_{N}(t) \right)\,.
\end{equation}
The above expression suggests to define the following new stochastic
processes:
\begin{eqnarray}
w(t) & = & \frac{1}{\sqrt{\lambda}} \sum_{n=1}^{N} \sqrt{\lambda_{n}}  w_{n}(t),
\label{eq:wnoise}\\
\tilde{w}_{n}(t) & = & w_{n}(t) - \frac{\lambda_{n}}{\lambda_{N}}\,
w_{N}(t), \qquad n=1,2, \ldots, N-1,
\end{eqnarray}
with $\lambda$ yet to be defined. The first noise is associated  to
the center-of-mass motion, the remaining ones are associated to the
relative motion. In such a way, Eq.~\eqref{eq:dfgf} becomes:
\begin{equation} \label{eq:xcg}
\sum_{n=1}^{N} \sqrt{\lambda_{n}}\, q_{n}\, w_{n}(t) \; = \;
\sqrt{\lambda}\, Q \, w(t) + \sum_{n=1}^{N-1} \sqrt{\lambda_{n}}\,
\tilde{q}_{n}\, \tilde{w}_{n}(t).
\end{equation}
The new noises here introduced are still Gaussian processes with zero
mean; a crucial role is of course played by their second moments. The
correlation function between $w(t)$ and $\tilde{w}_{n}(t)$ is
equal to:
\begin{equation}
{\mathbb E}_{\mathbb Q} [w(t)\, \tilde{w}_{n}(s)] \; = \;
\frac{1}{\sqrt{\lambda}}\, \sum_{m=1}^{N} \left[ \sqrt{\lambda_{m}}
D_{mn}(t,s) - \sqrt{\frac{\lambda_{n}\lambda_{m}}{\lambda_{N}}}\,
D_{mN}(t,s) \right], \qquad n=1,2, \ldots N-1.
\end{equation}
As we shall see soon, the {\it necessary condition} in order for the two
types of motion to decouple is that the above expression vanishes
\begin{equation} \label{eq:sdf}
\sum_{m=1}^{N} \left[ \sqrt{\lambda_{m}} D_{mn}(t,s) -
\sqrt{\frac{\lambda_{n}\lambda_{m}}{\lambda_{N}}}\, D_{mN}(t,s)
\right] \; = \; 0, \qquad n=1,2, \ldots N-1,
\end{equation}
i.e. that $w(t)$ be statistically independent from the other noises.
This condition is automatically satisfied if the original noises
$w_{n}(t)$ are {\it independent} and {\it identically} distributed, i.e. if
\begin{equation} \label{eq:iirv}
D_{nm}(t,s) \; = \; \delta_{nm}D(t,s),
\end{equation}
which is what it is usually assumed in white-noise collapse models
(see Eq.~\eqref{eq:lemp}). Throughout the rest of the  section we
will assume that this condition is satisfied.

The correlation function of $w(t)$ is equal to:
\begin{equation}
{\mathbb E}_{\mathbb Q} [w(t)\, w(s)] \; = \; \frac{1}{\lambda}\,
\left( \sum_{n=1}^{N} \lambda_{n} \right)\, D(t,s)\,;
\end{equation}
this expression suggests to set the value of the coupling
constant $\lambda$ first introduced in Eq.~\eqref{eq:wnoise} equal
to:
\begin{equation} \label{eq:ftghdf}
\lambda \; \equiv \; \sum_{n=1}^{N} \lambda_{n} \; = \;
\frac{M}{m_{0}}\, \lambda_{0},
\end{equation}
so that the correlation function of the center-of-mass noise is
identical to that of the noises associated to each single
constituent. Thus, in accordance with what happens in the white
noise case, the collapse coupling constant associated to the
center-of-mass motion is proportional to the total mass of the
system; as we shall see in the subsequent sections, this means  that
the collapse mechanism scales with the size of the system.

According to the previous assumptions, the correlation functions of the noises $\tilde{w}_{n}(t)$ are
equal to:
\begin{eqnarray}
{\mathbb E}_{\mathbb Q} [\tilde{w}_{n}(t)\, \tilde{w}_{m}(s)] & =
& C_{nm}\, D(t,s),\\
 C_{nm} & \equiv & \delta_{nm} \, + \,
\frac{\sqrt{\lambda_{n} \lambda_{m}}}{\lambda_{N}},
\end{eqnarray}
with $n,m = 1,2, \ldots, N-1$. It easy to prove that the matrix
$C$, having $C_{nm}$ as coefficients, is positive-definite. In fact
one can write $C=I-\mathbf{X}\mathbf{Y}^{\top}$, where
$\mathbf{X},\mathbf{Y}$ are two $N-1$-dimensional vectors,
$\mathbf{Y}=-\mathbf{X}$ and
\begin{equation}
\mathbf{X}= \frac{1}{\sqrt{\lambda_N}}\left(
\begin{array}{c}
\sqrt{\lambda_1}\\
\vdots\\
\sqrt{\lambda_{N-1}}
\end{array}
\right)\,.
\end{equation}
According to Statement 3.5.14 at page 93 in~\cite{Bernstein:05},
$C$ is positive definite iff $\mathbf{X}^{\top}\mathbf{Y}<1$,
inequality that can be easily verified with our definitions of
$\mathbf{X}$ and $\mathbf{Y}$:
\begin{equation}
\mathbf{X}^{\top}\mathbf{Y}=-\frac{1}{\lambda_N}
\sum_{i=1}^{N-1}\lambda_i=1-\frac{\lambda}{\lambda_N}<1\,.
\end{equation}

\noindent Moreover, $C$ is real and symmetric, thus there  exists an
orthogonal matrix $O$ which diagonalizes it:
\begin{equation}
O^{\top}\, C \, O \; = \; D,\qquad\qquad D=\mathrm{diag} (d_1,\dots d_{N-1})\,,
\end{equation}
where the eigenvalues $d_i$ of the matrix $D$ are real and
positive, since $C$ is positive definite.

It is convenient to  \lq\lq diagonalize'' the processes
$\tilde{w}_{1}(t),\dots,\tilde{w}_{N-1}(t)$ by defining the new
stochastic processes:
\begin{equation}\label{eq:wbar}
\overline{w}_{n}(t) \; \equiv \; \frac{1}{\sqrt{d_{n}}}\,
\sum_{m=1}^{N-1} O^{\top}_{nm}\, \tilde{w}_{m}(t);
\end{equation}
an easy calculation shows that they are independent and identically
distributed Gaussian processes with zero mean and correlation
function equal to $D(t,s)$. By defining also:
\begin{equation}\label{eq:lbar}
\overline{q}_{n} \; \equiv \; \frac{1}{\sqrt{\lambda_{n}}}
\sum_{m=1}^{N-1} O^{\top}_{nm} \, \sqrt{\lambda_{m}} \,
\tilde{q}_{m}, \qquad\quad \overline{\lambda}_{n} \; = \;
\lambda_{n} \, d^{2}_{n}\,,
\end{equation}
one can easily verify that the second term on the right hand side of
Eq.~\eqref{eq:xcg} can be rewritten as:
\begin{equation} \label{eq:xcgdf}
\sum_{n=1}^{N-1} \sqrt{\lambda_{n}}\, \tilde{q}_{n}\,
\tilde{w}_{n}(t) \; = \; \sum_{n=1}^{N-1} \sqrt{\bar{\lambda}_{n}}\,
\overline{q}_{n}\, \overline{w}_{n}(t)\,.
\end{equation}

To summarize, the second term on the right hand side of Eq.~\eqref{eq:main} becomes
\begin{equation}
\sum_{n=1}^{N} \sqrt{\lambda_{n}}\, q_{n}\,
w_{n}(t) \; =  \;\sqrt{\lambda}\,Q\, w(t)\;+\; \sum_{n=1}^{N-1} \sqrt{\bar{\lambda}_{n}}\,
\overline{q}_{n}\, \overline{w}_{n}(t)\,.
\end{equation}
We now examine the third term of Eq.~\eqref{eq:main}. After some easy
calculations, one finds the following result:
\begin{eqnarray}\label{eq:long}
\lefteqn{\sum_{n,m=1}^{N} \sqrt{\lambda_{n}} q_{n} \int_{0}^{t} ds
D_{nm}(t,s) \frac{\delta}{\delta w_{m}(s)}  = }\nonumber\\
&=& \sqrt{\lambda}\, Q
\int_{0}^{t} ds\, {\mathbb E}_{\mathbb Q} [w(t)\, w(s)]\,
\frac{\delta}{\delta w(s)}
 +  \sqrt{\lambda}\, Q \sum_{n=1}^{N-1} \int_{0}^{t} ds\, {\mathbb
E}_{\mathbb Q} [w(t)\, \tilde{w}_{n}(s)]\, \frac{\delta}{\delta
\tilde{w}_{n}(s)} \nonumber \\
& + & \sum_{n=1}^{N-1} \sqrt{\lambda_{n}}\, \tilde{q}_{n}
\int_{0}^{t} ds\, {\mathbb E}_{\mathbb Q} [\tilde{w}_{n}(t)\,
w(s)]\, \frac{\delta}{\delta
w(s)} + \sum_{n,m=1}^{N-1} \sqrt{\lambda_{n}}\, \tilde{q}_{n}
\int_{0}^{t} ds\, {\mathbb E}_{\mathbb Q} [\tilde{w}_{n}(t)\,
\tilde{w}_{m}(s)]\, \frac{\delta}{\delta \tilde{w}_{m}(s)}.\hspace{1cm}
\end{eqnarray}
As we see, the second and third term at the right hand side of the above
equation vanish when condition~\eqref{eq:sdf} is satisfied, in which
case the variables relative to the motion of the center of mass and
those relative to the motion of the relative degrees of freedom
decouple. This is the desired result. Moreover, under
condition~\eqref{eq:iirv} one can rewrite Eq.~\eqref{eq:long} as follows:
\begin{eqnarray}
\lefteqn{\sum_{n,m=1}^{N} \sqrt{\lambda_{n}} q_{n} \int_{0}^{t} ds
D_{nm}(t,s) \frac{\delta}{\delta w_{m}(s)}  =}\nonumber\\
&=& \sqrt{\lambda}\, Q
\int_{0}^{t} ds\, D(t,s)\, \frac{\delta}{\delta w(s)}
+ \sum_{n,m=1}^{N-1} \sqrt{\lambda_{n}}\, \tilde{q}_{n}
\int_{0}^{t} ds\, C_{nm} D(t,s)\, \frac{\delta}{\delta
\tilde{w}_{m}(s)} \nonumber \\
& = & \sqrt{\lambda}\, Q \int_{0}^{t} ds\, D(t,s)\,
\frac{\delta}{\delta w(s)} + \sum_{n}^{N-1}
\sqrt{\bar{\lambda}_{n}}\, \overline{q}_{n} \int_{0}^{t} ds\,
D(t,s)\, \frac{\delta}{\delta \overline{w}_{n}(s)}.
\end{eqnarray}

To summarize the results so far obtained, we have seen that when
condition~\eqref{eq:sdf} is satisfied, the center of mass motion and
the relative motion decouple; under the simplifying
assumption~\eqref{eq:iirv},  the equation for the center of mass
becomes:
\begin{equation}\label{eq:cm}
\frac{d}{dt}\, \phi_{t}^{\text{\tiny cm}} = \left[ -\frac{i}{\hbar}
H_{\text{\tiny cm}} + \sqrt{\lambda}\, Q\, w(t) - 2 \sqrt{\lambda} Q
\int_{0}^{t} ds D(t,s)
\frac{\delta}{\delta w(s)} \right] \phi_{t}^{\text{\tiny cm}},
\end{equation}
where $w(t)$ and $\lambda$ have been defined in~\eqref{eq:wnoise}
and~\eqref{eq:ftghdf} respectively. The equation for the relative
motion instead becomes:
\begin{equation} \label{eq:klfgf}
\frac{d}{dt}\, \phi_{t}^{\text{\tiny rel}} = \left[ -\frac{i}{\hbar}
H_{\text{\tiny rel}} + \sum_{n=1}^{N} \sqrt{\bar{\lambda}_{n}}
\bar{q}_{n} \bar{w}_{n}(t) - 2 \sum_{n,m=1}^{N-1}
\sqrt{\bar{\lambda}_{n}} \bar{q}_{n}C_{nm} \int_{0}^{t} ds D(t,s)
\frac{\delta}{\delta \bar{w}_{m}(s)} \right] \phi_{t}^{\text{\tiny
rel}},
\end{equation}
where $\bar{w}_n(t)$ and $\bar{\lambda}_n$ have been defined in~\eqref{eq:wbar}
and~\eqref{eq:lbar} respectively.

We stress once more a crucial feature of the model,  the
amplification mechanism: according to Eq.~\eqref{eq:ftghdf}, the
coupling between the noise and the center of mass, thus the collapse
strenght, scales with the size of the system. \noindent On the other
side, relative degrees of freedom are coupled to the noise through
$\lambda_n$, which do not scale with the size of the system, and
remain small. This is the way in which the model described by
Eq.~\eqref{eq:main} describes both classical properties of
macroscopic objects (large values of $m$) and quantum properties of
microscopic systems (small values of $m$).

\section{Solution of the equation for a free quantum particle}
\label{sec:three}

Now we restrict our attention to the dynamics of a free quantum particle
of mass $m$. Eq.~\eqref{eq:main} then becomes:
\begin{equation} \label{eq:freepart}
\frac{d}{dt}\, \phi_t(x) \; = \; \left[ - \frac{i}{\hbar}\,
\frac{p^2}{2m} \, + \, \sqrt{\lambda} q w(t) \, - \, 2
\sqrt{\lambda} q \int_0^t ds D(t,s) \frac{\delta}{\delta w(s)}
\right] \phi_t(x),
\end{equation}
where $D(t,s)$ is the correlation function of the Gaussian noise
$w(t)$. As discussed in the previous section, this same  equation
describes also the dynamics of the center of mass of an isolated
system (in particular, a macroscopic object) of total mass $m$.

Aim of this section is to find the explicit solution of
Eq.~\eqref{eq:freepart}.  In~\cite{Strunz:96} it has been shown that
the Green's function $G(x,t;x_0,0)$ of Eq.~\eqref{eq:freepart}
allows for the following path-integral representation:
\begin{equation}\label{eq:propform}
G(x,t;x_0,0) \; = \; \int^{q(t)=x}_{q(0)=x_0} \mathcal{D}[q] e^{\mathcal{S}[q]} \,,
\end{equation}
where, as indicated, the integration is carried out over all the
paths connecting $q(0)=x_0$ to $q(t)=x$. The `action'
$\mathcal{S}[q]$, which contrary to the standard quantum case,  has
both a real and an imaginary part, is:
\begin{equation} \label{eq:action}
\mathcal{S}[q] \; = \; \int^t_0 ds \left[ \frac{i m}{2\hbar}\,
q'^2(s) + \sqrt{\lambda} q(s) w(s) - \lambda q(s) \int_0^t dr\,
q(r) D(s,r)\right].
\end{equation}
That Eq.~\eqref{eq:action} represents the correct action associated to
Eq.~\eqref{eq:freepart} can be easily verified by checking that the
wave function
\begin{equation}\label{eq:general}
\phi_t(x) \; = \; \int_{-\infty}^{\infty} dx_0 G(x,t;x_0,0) \phi_0(x_0)\,,
\end{equation}
solves Eq.~\eqref{eq:freepart}. The advantage of the path integral
representation of the Green's function, with respect to the standard
representation associated to a differential equation
like~\eqref{eq:freepart}, is that it avoids resorting to the
functional derivative of the noise, which is a source of major
complications. We now compute the path
integration in~\eqref{eq:propform}.

Following the standard Feynman polygonal approach~\cite{Feynman:65}, we
divide the time interval $[0,t]$ in $N$ subintervals, each of length
$\epsilon = t/N$; the intermediate time points are defined as: $t_k
= k \epsilon$. The path integral is then understood as the limit
$N\rightarrow\infty$ of a multiple integral over the $N-1$ variables
$q_k = q(t_k), k = 1, \ldots, N-1$:
\begin{equation}
G(x,t;x_0,0) = \lim_{N \rightarrow \infty} G_N(x,t;x_0,0)\, ,
\end{equation}
with
\begin{equation}
G_N(x,t;x_0,0) = \left(\frac{m}{2\pi
i\hbar\epsilon}\right)^{\frac{N}{2}}\int\cdots\int\prod_{k=1}^{N-1} dq_k
\,e^{\mathcal{S}_N[q]}\,,
\end{equation}
and $\mathcal{S}_N[q]$ is the discretized form of the `action',
which reads:
\begin{equation}
\mathcal{S}_N[q]= \sum_{k=1}^N\left[\frac{i}{\hbar}
\frac{m}{2\epsilon}(q_k-q_{k-1})^2 +
\epsilon\sqrt{\lambda}  w_k q_k - \epsilon^2 \lambda
q_k\sum_{j=1}^N D_{k,j}q_j\right],
\end{equation}
where $w_k = w(t_k)$ and $D_{k,j}=D(t_k,t_j)$. The constraints are:
$q_0 = x_0$ and $q_N = x$. We re-write $G_N(x,t;x_0,0)$, separating
the terms which are constant with respect to the integration
variables, the linear terms and the quadratic terms; using a vector
notation, we have:
\begin{eqnarray}\label{eq:propdisc}
G_N(x,t;x_0,0) & = & \left(\frac{m}{2\pi
i\hbar\epsilon}\right)^{\frac{N}{2}}\exp
\left[\frac{im}{2\hbar\epsilon}\left(x_0^2+x^2\right) + \epsilon
\sqrt{\lambda}\, w_N x
-\epsilon^2\lambda x^2 D_{N,N} \right]\hspace{2cm}\nonumber \\
& & \hspace{5cm}\cdot\int_{-\infty}^{+\infty}d\mathbf{X} \exp\left[-\mathbf{X}
\cdot A \mathbf{X} + 2 \, \mathbf{X} \cdot \mathbf{Y}\right]\,,
\end{eqnarray}
where $\mathbf{X}$ and $\mathbf{Y}$ are two $(N-1)$-dimensional
vectors defined as follows:
\begin{equation}\label{eq:xy}
\mathbf{X}= \left(
\begin{array}{c}
q_1\\
q_2\\
\vdots\\
q_{N-2}\\
q_{N-1}
\end{array}
\right)\, ,\qquad
\mathbf{Y}= -\frac{im}{2\hbar\epsilon}\left(
\begin{array}{c}
x_0\\
0\\
\vdots\\
0\\
x
\end{array}
\right) +\frac{\epsilon \sqrt{\lambda}}{2}\left(
\begin{array}{c}
w_1\\
w_2\\
\vdots\\
w_{N-2}\\
w_{N-1}
\end{array}
\right) -\frac{\epsilon^2\lambda}{2}\, x\left(
\begin{array}{c}
D_{1, N}\\
D_{2, N}\\
\vdots\\
D_{N-2, N}\\
D_{N-1, N}
\end{array}
\right) \,.
\end{equation}
We have also used the short hand notation:
$d\mathbf{X}=\prod_{k=1}^{N-1} dq_k$. The matrix $A$ is the sum of
two symmetric, $(N-1)$-dimensional square matrices $B$ and $C$,
whose entries are:
\begin{eqnarray}
\label{eq:b} B_{i,i}&=&-\frac{im}{\hbar\epsilon},\quad\quad B_{i,i\pm 1} =
\frac{im}{2\hbar\epsilon},\\
 B_{i,j}&=&0\,, \;\; j\neq i,i\pm1\,, \label{eq:rty1}\\
\label{eq:c} C_{i,j}&=&\epsilon^2\lambda D_{i,j}\,.  \label{eq:rty2}
\end{eqnarray}
The multiple Gaussian integral of Eq.~\eqref{eq:propdisc} can be
immediately evaluated by using the standard result:
\begin{equation}\label{eq:gaussint}
\int_{-\infty}^{+\infty}d\mathbf{X}\,\exp\left[-\mathbf{X} \cdot A
\mathbf{X} + a \mathbf{X} \cdot \mathbf{Y}\right] =
\sqrt{\frac{\pi^{N-1}}{\det(A)}} \,\exp\left[\frac{a^2}{4}
\mathbf{Y} \cdot A^{-1}\mathbf{Y}\right];
\end{equation}
$G_N(x,t;x_0,0)$ then becomes:
\begin{equation}\label{eq:propdisc2}
G_N(x,t;x_0,0)= \sqrt{\frac{(m/2 i\hbar\epsilon)^{N}}{\pi\det(A)}}
\exp\left[ \mathbf{Y} \cdot A^{-1}\mathbf{Y}
+\frac{im}{2\hbar\epsilon} \left(x_0^2+x^2\right)+ \epsilon
\sqrt{\lambda}\, w_N x -\epsilon^2\lambda x^2 D_{N,N}\right].
\end{equation}

Note that the integral in Eq.~\eqref{eq:gaussint} exists only if $A$ is a positive definite matrix. The result can be extended to non-negative matrices, following e.g. the procedure of Theorem 1, page 13 of~\cite{Gikhman:96}. Accordingly, our results strictly hold only when the correlation function $D(t,s)$ of the noise $w(t)$, seen as an integral kernel, is non-negative definite.

The next step is to take the limit $N \rightarrow \infty$, in which
case one encounters two main difficulties: the first is to evaluate
the inverse matrix $A^{-1}$, the second is to compute the
determinant of $A$, in both cases for any $N$. To solve these
difficulties, we proceed as in~\cite{Khandekar:83}.

In order to evaluate $\mathbf{Y} \cdot A^{-1}\mathbf{Y}$, we
introduce a twice differentiable function $z(s), s \in [0,t]$, yet
to be determined, such that, given the vector $\mathbf{Z} := (z_1,
\ldots z_{N-1})^{\top}$, with $z_k = z(t_k), k = 1, \ldots N-1$, the
following matrix equation is satisfied:
\begin{equation} \label{eq:matr}
A\mathbf{Z}=\mathbf{Y}\,.
\end{equation}
For reasons which will be clear soon, we will set $z(0) = x_0$
and $z(t) = x$. We then have:
\begin{eqnarray}
\mathbf{Y}\cdot A^{-1}\mathbf{Y}&=&\mathbf{Y}\cdot\mathbf{Z}\nonumber\\
& = & -\frac{im}{2\hbar\epsilon}\left(x_0z_1 + x\, z_{N-1}\right) +
\frac{\epsilon\sqrt{\lambda}}{2}\sum_{j=1}^{N-1}w_jz_j-
\frac{\epsilon^2\lambda}{2}x\sum_{j=1}^{N-1}D_{j,N}z_j \nonumber \\
&=&-\frac{im}{2\hbar\epsilon}x_0\left(z(0)+\epsilon z'(0)+
O(\epsilon^2)\right)-\frac{im}{2\hbar\epsilon}x\left(z(t)-
\epsilon z'(t)+O(\epsilon^2)\right)+\nonumber\\
& & + \frac{\epsilon\sqrt{\lambda}}{2}\sum_{j=1}^{N-1}
w_jz_j-\frac{\epsilon^2\lambda}{2}x\sum_{j=1}^{N-1}D_{j,N}z_j;
\label{eq:dfgm}
\end{eqnarray}
the prime in $z'(s)$ denotes the first derivative of $z(s)$.
Inserting the above result in~\eqref{eq:propdisc2}, we have:
\begin{equation}\label{eq:propdisc3}
G_N(x,t;x_0,0)= \sqrt{\frac{(m/2 i\hbar\epsilon)^{N}}{\pi\det(A)}}
\exp\left[-\frac{im}{2\hbar}\left(x_0 z'(0) - x z'(t)\right) +
\frac{\epsilon\sqrt{\lambda}}{2}
\sum_{j=1}^{N-1}w_jz_j+O(\epsilon)\right],
\end{equation}
where we have collected all terms of order $\epsilon$ or higher,
which will vanish in the limit $N \rightarrow \infty$. Note that the
two terms of~\eqref{eq:propdisc2} proportional to $x_0^2$ and $x^2$,
which are of order $\epsilon^{-1}$, are canceled by the analogous
terms in~\eqref{eq:dfgm}: this is a consequence of the boundary
conditions we have set on the function $z(s)$.

In order to compute the determinant of the matrix $A=B+C$,  where
$B$ and $C$ have been defined in~\eqref{eq:b} and~\eqref{eq:c}, it
is convenient to isolate the determinant of $B$ by writing
$A=B(I+B^{-1}C)$, so that:
\begin{equation}
\det(A)=\det(B)\det(I+B^{-1}C)\, .
\end{equation}
The quantity $\det(B)$ will be used to simplify part of the terms
under the square root in~\eqref{eq:propdisc3}. The matrix $B$ is
equal to $B = \frac{m}{2i\hbar \epsilon}\, \overline{B}$ with
\begin{equation}
\overline{B}_{i,i} = 2, \qquad \overline{B}_{i,i\pm1} = -1,
\quad \overline{B}_{i,j} = 0, \;\; j \neq i, i \pm1.
\end{equation}
Let us call $\Delta_k$ the determinant of the matrix obtained from
$\overline{B}$ by removing the first $N-k-1$ rows and columns; the
following recursive relation is easy to prove:
\begin{equation}
\Delta_k=2\Delta_{k-1}-\Delta_{k-2},\qquad k = 1, \ldots N-1,
\end{equation}
where we have set $\Delta_0=1$ and $\Delta_{-1}=0$. From this
relation, by induction one immediately sees that $\Delta_k = k+1$,
which means that $\det(\overline{B}) = N$. Accordingly we have:
\begin{equation}
\det(B)=N \left(\frac{m}{2i\hbar\epsilon}\right)^{N-1}\,.
\end{equation}
Collecting all results we can write:
\begin{equation}\label{eq:propdisc4}
G_N(x,t;x_0,0)= \sqrt{\frac{m}{2i\pi\hbar\, t\, u_N(t)}}
\exp\left[-\frac{im}{2\hbar}\left(x_0 z'(0) - x z'(t)\right) +
\frac{\epsilon\sqrt{\lambda}}{2}
\sum_{j=1}^{N-1}w_jz_j+O(\epsilon)\right],
\end{equation}
where $t = N \epsilon$, and we have introduced the quantity $u_N(t)
: = \det(I+B^{-1}C)$. We are now in the position to perform the
limit $N \rightarrow \infty$; the Green's function becomes:
\begin{equation}\label{eq:prop}
G(x,t;x_0,0)= \sqrt{\frac{m}{2i\pi\hbar\, t\, u(t)}}
\exp\left[-\frac{im}{2\hbar}\left(x_0 z'(0) - x z'(t)\right)
+\frac{\sqrt{\lambda}}{2}\int_0^t ds\, w(s)z(s)\right],
\end{equation}
which is defined in terms of the two unknown functions $z(s)$ and
$u(t) := \lim_{N \rightarrow \infty} u_N(t)$. We now show how to
determine them.

Written in components and once divided by $\epsilon$,
Eq.~\eqref{eq:matr} becomes:
\begin{equation}
\frac{im}{2\hbar\epsilon^2}\left(z_{k+1}-2z_k+z_{k-1}\right)+
\epsilon\lambda \sum_{j=1}^{N-1}D_{k,j}z_j =
\frac{\sqrt{\lambda}}{2}w_k-\frac{\epsilon\lambda}{2} x D_{k,N},
\qquad k = 1, \ldots N-1;
\end{equation}
note that for $k = 1$ and for $k = N-1$, consistency is assured
thanks to the assumption $z_0 = x_0$ and $z_N = x$. Taking the limit
$\epsilon\rightarrow 0$ ($N\rightarrow\infty$) we obtain:
\begin{equation}\label{eq:z}
\frac{im}{2\hbar}\, z''(s)+\lambda\int_0^t dr\, D(s,r)z(r) \; = \;
\frac{\sqrt{\lambda}}{2}w(s),
\end{equation}
which is an integro-differential equation to be solved together with
the conditions $z(0)=x_0$, $z(t)=x$. The above equation determines
$z(s)$. In appendix~\ref{sec:app} we prove that Eq.~\eqref{eq:z} admits a unique solution.

Let us now determine $u(t)$; we follow once again the procedure
outlined in~\cite{Khandekar:83}. According to Statement 11.11.4, page 451 of~\cite{Bernstein:05} one can write
\begin{equation}
\det(I-\eta K)=\exp\left[\int_{\eta}^0\mathrm{Tr}[R(\mu)]d\mu\right]\,,
\end{equation}
where $R(\mu)$ is a matrix satisfying the equation $R(\mu)=K+\mu KR$. Applied to our problem, we have:
\begin{equation}\label{eq:detn}
u_N(t) \; = \; \det(I+B^{-1}C) \; = \; \exp\left[\int_{-1}^0\left(
\epsilon \sum_{k=1}^{N-1} R_{k,k}(\mu)\right)d\mu \right]\, ;
\end{equation}
in our case $K = B^{-1}C$, thus $R(\mu)$ satisfies the matrix
equation $B R(\mu) = C\epsilon^{-1} + \mu C R(\mu)$. Written in
components, and divided by $\epsilon$, this equation becomes:
\begin{equation}
\frac{im}{2 \hbar \epsilon^2} \left[ R_{i-1,j}(\mu) - 2R_{i,j}(\mu) +
R_{i+1,j}(\mu) \right] \; = \; \lambda D_{i,j} + \mu\, \epsilon
\sum_{k=1}^{N-1}D_{i,k}R_{k,j}(\mu),
\end{equation}
with the boundary conditions $R_{0,j}(\mu) = R_{N,j}(\mu) = 0$ for any
$j$ and $\mu$. Taking the limit $\epsilon \rightarrow 0$, the
matrix $R_{i,j}(\mu)$ becomes a function $R(s,r,\mu)$ satisfying the
following integro-differential equation:
\begin{equation}\label{eq:R}
\frac{im}{2\hbar}\frac{\partial^2}{\partial s^2} R(s,r,\mu) -
\lambda \mu \int_0^t dl\, D(s,l)R(l,r,\mu) \; = \; \lambda D(s,r),
\end{equation}
which has to be solved together with the boundary conditions
$R(0,r,\mu) = R(t,r,\mu) = 0$ for any $r \in [0,t]$ and $\mu \in
[-1,0]$. The existence and uniqueness theorem of appendix~\ref{sec:app} applies also to this equation. In the limit $N \rightarrow \infty$,
Eq.~\eqref{eq:detn} becomes:
\begin{equation}\label{eq:det}
u(t)  \; = \; \exp\left[\int_{-1}^0 d\mu\left( \int_0^t ds\,
R(s,s,\mu)\right) \right].
\end{equation}

\noindent Eq.~\eqref{eq:prop}, with $z(s)$ defined by~\eqref{eq:z} and $u(t)$
defined by~\eqref{eq:det}, gives a complete description of the
Green's function of Eq.~\eqref{eq:freepart}. This is the main result
of this section.

The function $z(s)$ solution of Eq.~\eqref{eq:z}  depends on the end
points $x_0$ and $x$ because of the boundary conditions. This is not
evident in the expression~\eqref{eq:prop} for the propagator. In
order to make such a dependence on $x_0$ and $x$ explicit, we
rewrite $z(s)$ as follows:
\begin{equation}\label{eq:ans}
z(s)=f(s)x_0+g(s)x+h(s)\,,
\end{equation}
where $f(s)$ satisfies the  homogenous integro-differential
equation:
\begin{equation}\label{eq:fans}
\frac{im}{2\hbar}\, f''(s)+\lambda\int_0^t dr\, D(s,r)f(r) \; = \;0\,,
\end{equation}
with boundary conditions  $f(0)=1$, $f(t)=0$, while $g(s)$ solves the same equation but with boundary conditions $g(0)=0$, $g(t)=1$.
The function $h(s)$ instead satisfies the non-homogenous equation:
\begin{equation}\label{eq:hans}
\frac{im}{2\hbar}\, h''(s)+\lambda\int_0^t dr\, D(s,r)h(r) \; = \;
\frac{\sqrt{\lambda}}{2}w(s),
\end{equation}
with boundary conditions $h(0)=h(t)=0$.
Note that $f(s)$, $g(s)$ and $h(s)$ depend also in the parameter $t$, so one should more properly write $f_t(s)$, $g_t(s)$, $h_t(s)$; sometimes we will omitt the pedex $t$ when no confusion arises. Here and in the following, the prime denotes differentiation with respect to the variable in parentheses.

With the help of~\eqref{eq:ans} one can write the Green's
function of Eq.~\eqref{eq:prop} in the following way:
\begin{equation}\label{eq:propexp}
G(x,t;x_0,0) = \sqrt{\frac{m}{2i\pi\hbar\, t\,
u(t)}}\exp\left[-\mathcal{A}_t
x_0^2-\tilde{\mathcal{A}}_tx^2+\mathcal{B}_tx_0x
+\mathcal{C}_tx_0+\mathcal{D}_tx +\mathcal{E}_t\right]\,,
\end{equation}
where the dependence on $x_0$ and $x$ has been made explicit. The
coefficients $\mathcal{A}_t$, $\tilde{\mathcal{A}}_t$ and
$\mathcal{B}_t$ are deterministic functions of time, defined as
follows:
\begin{equation}\label{eq:matha}
\mathcal{A}_t=kf_t'(0)\,, \qquad\tilde{\mathcal{A}}_t=-kg_t'(t)\,,
\qquad\mathcal{B}_t=k(f_t'(t)-g_t'(0)), \qquad\qquad
k=\frac{im}{2\hbar}\,,
\end{equation}
while the coefficients $\mathcal{C}_t$, $\mathcal{D}_t$ and
$\mathcal{E}_t$ are random processes:
\begin{eqnarray}
\mathcal{C}_t&=&-kh_t'(0) +\frac{\sqrt{\lambda}}{2}\int_0^tw(l)f_t(l)dl\,,\\
\mathcal{D}_t&=&kh_t'(t)+\frac{\sqrt{\lambda}}{2}\int_0^tw(l)g_t(l)dl\\
\label{eq:mathe}\mathcal{E}_t&=&\frac{\sqrt{\lambda}}{2}\int_0^tw(l)h_t(l)dl\,.
\end{eqnarray}

The  propagator~\eqref{eq:propexp} takes a simpler and more
symmetric form when the noise is time translation invariant:
$D(t,s)=D(t-s)$; according to the symmetry in the $(t,s)$ variables
one also has: $D(t-s)=D(|t-s|)$. In this case, one can easily verify
that if $f(s)$ is solution of Eq.~\eqref{eq:fans} with the boundary
conditions $f(0) = 1$, $f(t) = 0$, then $\tilde{f}(s)=f(t-s)$ is
still a solution Eq.~\eqref{eq:fans}, with boundary conditions
$\tilde{f}(0)=0$ and $\tilde{f}(t)=1$; in other words:
$\tilde{f}(s)=g(s)$.
Then~\eqref{eq:ans} becomes:
\begin{equation}\label{eq:ansnogh}
z(s) = f(s)x_0 + f(t-s)x -h(s)\,,
\end{equation}
and the Green's function~\eqref{eq:propexp} simplifies as
follows:
\begin{equation}\label{eq:propts}
G(x,t;x_0,0)=\sqrt{\frac{m}{2i\pi\hbar\, t\, u(t)}}
\exp\left[-\mathcal{A}_t (x_0^2+x^2) + \mathcal{B}_tx_0x +
\mathcal{C}_tx_0+\mathcal{D}_tx+\mathcal{E}_t\right]\,,
\end{equation}
where the deterministic coefficients now are:
\begin{equation} \label{eq:gvxshnj}
\mathcal{A}_t=kf_t'(0)\,,\qquad\mathcal{B}_t=2kf_t'(t)\,,
\end{equation}
and the stochastic terms read:
\begin{eqnarray}
\mathcal{C}_t & = & -k h_t'(0) + \frac{\sqrt{\lambda}}{2}\int_0^t dl\,
w(l)f_t(l),\\
\mathcal{D}_t & = &  k h_t'(t) + \frac{\sqrt{\lambda}}{2}\int_0^t dl\, w(l) f_t(t-l), \\
\mathcal{E}_t & = & \frac{\sqrt{\lambda}}{2}\int_0^t dl\, w(l) h_t(l)
\,.\label{eq:gvxshnj2}
\end{eqnarray}
Accordingly, for a general time translation invariant noise, the
structure of the Green's function is the same as in the familiar
white-noise case~\cite{Grosche:98,Bassi:08}, as we will review in
Sec.~\ref{sec:five}. Note that due to linearity, the general
solution of Eq.~\eqref{eq:hans} can be written as $h(s) =
h^{\text{\tiny P}}(s) + h^0(s)$, where $h^0(s)$ solves
the homogeneous equation ~\eqref{eq:fans} and $h^{\text{\tiny
P}}(s)$ is a particular solution of Eq.~\eqref{eq:hans}. Often the
particular solution admits an integral expression of the form:
\begin{equation} \label{eq:dfgdsfp}
h^{\text{\tiny P}}(s) \; = \; \int_0^s dl \, r(s,l),
\end{equation}
where $r(s,l)$ is some function. Eq.~\eqref{eq:dfgdsfp} implies that
$h^{\text{\tiny P}}(0) = 0$. One can immediately verify that, by
taking $h^0(s) = - h^{\text{\tiny P}}(t) f(t-s)$, the function:
\begin{equation} \label{eq:gfhsd}
h(s) = h^{\text{\tiny P}}(s) - h^{\text{\tiny P}}(t) f(t-s)\,,
\end{equation}
solves Eq.~\eqref{eq:hans}, with the correct boundary conditions.
Thus also $h(s)$ can be expressed in terms of $f(s)$, once
$h^{\text{\tiny P}}(s)$ is given as in~\eqref{eq:dfgdsfp}.

A concluding remark is at order. In this paper we have computed
the Green's function $G(x,t;x_0,0)$ by exploiting the path-integral
formalism. The same result can be obtained also by resorting to the
standard operator formalism, starting with the expression for the
propagator first proposed in~\cite{Pearle:96,Pearle:96bis}.

\section{The non Markovian stochastic Schr\"{o}dinger equation}
\label{sec:four}

The result of the previous section allows us to rewrite
Eq.~\eqref{eq:freepart} in a simpler form, removing the functional
derivative which is not easy to handle. In this way, we will
prove the correctness of the Ansatz introduced in Sec. 4.3 of Ref.~\cite{Diosi:98}, according to which the functional derivative can be replaced
with a suitable function of the position and momentum operators.

The starting point is  Eq.~\eqref{eq:general} where the solution
$\phi_t(x)$ of  Eq.~\eqref{eq:freepart} is expressed in terms of the
Green's function $G(x,t;x_0,0)$. Using standard path-integral
techniques, e.g. generalizing the procedure described at page 509
of~\cite{Muller:06}, one can show that $G(x,t;x_0,0)$ given
by~\eqref{eq:propform}, with $\mathcal{S}[q]$ given
by~\eqref{eq:action}, satisfies the following equation:
\begin{equation}\label{eq:blabla}
\frac{d}{dt}G(x,t;x_0,0)=\left[ -\frac{i}{\hbar}\, \frac{p^2}{2m} +
\sqrt{\lambda} x w(t)\right] G(x,t;x_0,0)+\tilde{G}(x,t;x_0,0)\,,
\end{equation}
with:
\begin{equation}\label{eq:gtilde}
\tilde{G}(x,t;x_0,0):=-2\lambda x\int^{q(t)=x}_{q(0)=x_0}
\mathcal{D}[q]\int_0^t ds\, q(s) D(t,s) e^{\mathcal{S}[q]}\,.
\end{equation}
The path integral in~\eqref{eq:gtilde} cannot be trivially reduced
to~\eqref{eq:propform} because $q(s)$ appears, which depends on the
entire time interval $[0,t]$. Thus we have to re-calculate
$\tilde{G}(x,t;x_0,0)$ as we did with $G(x,t;x_0,0)$. Proceeding as
in Section~\ref{sec:three}, we define
$\tilde{G}(x,t;x_0,0)=\lim_{N\rightarrow\infty}\tilde{G}_N(x,t;x_0,0)$,
where $\tilde{G}_N(x,t;x_0,0)$ is:
\begin{eqnarray}
\tilde{G}_N(x,t;x_0,0) & = &-2\lambda q \left(\frac{m}{2\pi
i\hbar\epsilon}\right)^{\frac{N}{2}}\exp
\left[\frac{im}{2\hbar\epsilon}\left(x_0^2+x^2\right) + \epsilon
\sqrt{\lambda}\, w_N x
-\epsilon^2\lambda x^2 D_{N,N} \right]\nonumber \\
& & \int_{-\infty}^{+\infty}d\mathbf{X}\,
\left(\epsilon\sum_{k=1}^{N-1}\mathbf{K}^{\top}
\mathbf{X}+O(\epsilon)\right) \exp\left[-\mathbf{X} \cdot A
\mathbf{X} + 2 \, \mathbf{X} \cdot \mathbf{Y}\right]\,,
\end{eqnarray}
where $O(\epsilon)$ collects all terms of order $\epsilon$,
$\mathbf{K}^{\top}=(0,\dots D_{Nk},\dots 0)$ where $D_{Nk}$ is placed in the
$k$-th position, and $\mathbf{X}$, $\mathbf{Y}$, $A$ are defined in Eqs.~\eqref{eq:xy} and~\eqref{eq:b}-\eqref{eq:c}.
Applying the following generalization of Eq.~\eqref{eq:gaussint}:
\begin{equation}\label{eq:gaussintgen}
\int_{-\infty}^{+\infty}d\mathbf{X}\,\mathbf{K}^{\top}\mathbf{X}\,
\exp\left[-\mathbf{X} \cdot A \mathbf{X} + a \mathbf{X} \cdot
\mathbf{Y}\right] = \sqrt{\frac{\pi^{N-1}}{\det(A)}}
\,\frac{a}{2}\mathbf{K}^{\top}A^{-1}\mathbf{Y}
\,\exp\left[\frac{a^2}{4} \mathbf{Y} \cdot
A^{-1}\mathbf{Y}\right]\,,
\end{equation}
and taking the limit for $\epsilon\rightarrow 0$ one finds that
\begin{equation}\label{eq:bla2}
\tilde{G}(x,t;x_0,0)=-2\lambda x\int_0^t ds\, z(s) D(t,s)G(x,t;x_0,0)\,,
\end{equation}
where $z(s)$ is defined in Eq.~\eqref{eq:z}. Thus,
using~\eqref{eq:blabla} and~\eqref{eq:bla2} one can formally write
($x=q$, the position operator):
\begin{equation}\label{eq:dg}
\frac{d}{dt}G(x,t;x_0,0)=\left[ -\frac{i}{\hbar}\,
\frac{p^2}{2m} + \sqrt{\lambda} q w(t) - 2\lambda q \int_0^t ds\,
z(s) D(t,s)\right] G(x,t;x_0,0)\,.
\end{equation}
Note that it is possible to find this same result by  inserting
$G(x,t;x_0,0)$ directly into Eq.~\eqref{eq:freepart}. However, to
reach the result of Eq.~\eqref{eq:dg}, one encounters a term of the
form $\delta z'(l)/\delta w(s)$ which can be simplified only under
the assumption that the time derivative and the functional
derivative acting on $z(s)$ commute. This is not true in general,
and one has to check it by finding the explicit functional
dependence of $z(s)$ on $w(s)$. This fact makes this second approach
more cumbersome than the one we have followed.

Eq.~\eqref{eq:dg} is not written in a useful way yet, since $z(s)$
depends both on $x_0$ and $x$. We make explicit this dependence on
$x_0$ and $x=q$ by resorting to~\eqref{eq:ans}:
\begin{equation}\label{eq:dg2}
\tilde{G}(x,t;x_0,0)= - 2\lambda x\left[  \int_0^t ds\, D(t,s) \big(f(s)x_0
+ g(s)x + h(s)\big)\right] G(x,t;x_0,0)\,.
\end{equation}
The term in~\eqref{eq:dg2} depending on $x_0$ does not allow to
perform the integration over the $x_0$ variable, to compute $\phi_t$
from~\eqref{eq:general}. This term can be quite easily written in
terms of the position operator $q$ and the momentum operator
$p=-i\hbar\frac{\partial}{\partial x}$ acting on $G(x,t;x_0,0)$, as
follows:
\begin{eqnarray}
x_0G(x,t;x_0,0)&=&\frac{1}{g'(0)-f'(t)}\Bigg[\frac{2i\hbar}{m}
\frac{\partial}{\partial
x}G(x,t;x_0,0)\hspace{1.0cm}\nonumber\\
&&\hspace{1.0cm}+\left(2xg'(t)+h'(t)+\frac{i\sqrt{\lambda}\hbar}{m}
\int_0^tdl\,w(l)g(l)\right)G(x,t;x_0,0)\Bigg]\,.\nonumber
\end{eqnarray}
Collecting all the above results, we then find that $\phi_t(x)$
satisfies the following equation:
\begin{equation} \label{eq:yhdfsd}
\frac{d}{dt}\phi_t(x)=\left[ -\frac{i}{\hbar}\,
\frac{p^2}{2m} + \sqrt{\lambda} q w(t)- 2\lambda q \int_0^t ds\,
D(t,s)\big(q\,a_t(s)+p\,b_t(s)+c_t(s)\big)\right]\phi_t(x)\,,
\end{equation}
where
\begin{eqnarray}
a_t(s)&=&g(s)+2\frac{g'(t)}{g'(0)-f'(t)}f(s)\,,\label{eq:xvhkg0}\\
b_t(s)&=&-\frac{2}{m}\frac{f(s)}{g'(0)-f'(t)}\,, \label{eq:xvhkg1}\\
c_t(s)&=&h(s)+\frac{f(s)}{g'(0)-f'(t)}\left(h'(t)+
\frac{i\sqrt{\lambda}\hbar}{m}\int_0^tdl\,w(l)g(l)\right)\,. \label{eq:xvhkg2}
\end{eqnarray}
This is the non-Markovian stochastic Schr\"odinger equation,  whose
Green's function is given by Eq.~\eqref{eq:propexp}.  When the noise
is time-translation invariant ($D(t,s)=D(|t-s|)$), the functions
$a_t(s)$, $b_t(s)$, $c_t(s)$ simplify as follows:
\begin{eqnarray}
a_t(s) & = & f(t-s) + \frac{f'(0)}{f'(t)}f(s)\,,\\
b_t(s) & =& \frac{1}{m}\frac{f(s)}{f'(t)} \,,\\
c_t(s) & = & h(s) -
\frac{f(s)}{2 f'(t)}\left(h'(t)+\frac{i\sqrt{\lambda}\hbar}{m}\int_0^tdl\,w(l)f(t-l)\right)\,.
\end{eqnarray}
As anticipated, Eq.~\eqref{eq:yhdfsd} agrees with the ansatz
first proposed in~\cite{Diosi:98}.

\section{White noise limit of the Green's function}
\label{sec:five}

Having found the Green's function associated to
Eq.~\eqref{eq:freepart}, the first task is to check that, in the
white noise limit, it coincides with the propagator already known in
the literature~\cite{Grosche:98,Bassi:08}. Since $D(s,r) =
\delta(s-r)$ is time translation invariant, we can  use the expression~\eqref{eq:ansnogh} for $z(s)$. In the white noise case,
Eq.~\eqref{eq:fans} reduces to the second order differential
equation:
\begin{equation}
f''(s) - \frac{2i\lambda\hbar}{m} f(s)=0\,, \qquad\qquad
f(0)=1,\quad f(t)=0\,,
\end{equation}
whose solution can be easily computed:
\begin{eqnarray}\label{eq:fwhite}
f(s)&=& \cosh\upsilon s  -\coth\upsilon t\,\sinh\upsilon s\,,\\
\upsilon & :=& \frac{1+i}{2}\, \omega, \qquad \omega
\; := \; 2 \sqrt{\frac{\hbar \lambda}{m}}\,.
\end{eqnarray}
Eq.~\eqref{eq:hans} for $h(s)$ instead reduces to:
\begin{equation}
h''(s) - \frac{2i\lambda\hbar}{m} h(s)= -
\frac{i\hbar}{m}\sqrt{\lambda}\, w(s), \qquad\qquad h(0)=h(t)=0\,.
\end{equation}
Using once again time translation invariance, we know that $h(s)$
takes the form~\eqref{eq:gfhsd}, where a particular solution
$h^{\text{\tiny P}}(s)$ is~\cite{Polyanin:03}:
\begin{equation}
h^{\text{\tiny P}}(s) =
-\frac{i\hbar}{m\upsilon}\sqrt{\lambda}\int_0^s dr\,
w(r)\sinh\upsilon(s-r).
\end{equation}
The Green's function then has the form~\eqref{eq:propts}, where the
deterministic coefficients $\mathcal{A}_t$ and $\mathcal{B}_t$ are
equal to:
\begin{equation}
\mathcal{A}_t \; = \; \frac{\lambda}{\upsilon}\coth\upsilon
t\,,\qquad\qquad
\mathcal{B}_t=\frac{2\lambda}{\upsilon}\sinh^{-1}\upsilon t,
\end{equation}
while the random coefficients $\mathcal{C}_t$, $\mathcal{D}_t$ and
$\mathcal{E}_t$ are equal to:
\begin{eqnarray}
\mathcal{C}_t&=&\sqrt{\lambda}\int_0^t dl\, w(l)
\frac{\sinh\upsilon(t-l)}{\sinh\upsilon t}\,,\\
\mathcal{D}_t&=&\sqrt{\lambda}\int_0^t dl\, w(l)
\frac{\sinh\upsilon l}{\sinh\upsilon t},\\
\mathcal{E}_t&=&\frac{\upsilon}{4}\left( \int_0^t dl\, w(l)
\sinh\upsilon l \int_0^t ds\, w(s)
\frac{\sinh\upsilon(t-s)}{\sinh\upsilon t}\right.\nonumber\\
&&\left.-  \int_0^t dl \,
w(l) \int_0^l ds\, w(s) \sinh\upsilon(l-s)\right)\,.\qquad
\end{eqnarray}
After some manipulation and an integration by parts, one can rewrite
$\mathcal{E}_t$ as follows:
\begin{eqnarray}
\label{eq:e1}\mathcal{E}_t&=&\frac{\upsilon}{2} \int_0^t dl \, w(l)
\frac{\sinh\upsilon(t-l)}{\sinh\upsilon t} \int_0^l ds\, w(s) \sinh\upsilon s\\
\label{eq:e2}&=&\frac{\upsilon^2}{4} \int_0^t dl \,
\frac{ \left(\int_0^l ds\, w(s) \sinh\upsilon s\right)^2}{\sinh^2\upsilon l}\\
&=&\frac{\upsilon^2}{4\lambda}\int_0^tdl\,\mathcal{D}_l\,.
\end{eqnarray}
When $\mathcal{E}_t$ is written as in Eq.~\eqref{eq:e1}, then the
Green's function~\eqref{eq:prop} reduces to that of Eq.~6.2.42, page
180 of~\cite{Grosche:98}. When $\mathcal{E}_t$ is rewritten as in
Eq.~\eqref{eq:e2}, it coincides with the corresponding term
of~\cite{Bassi:08}.

We now compute $R(s,r,\mu)$, which in turn determines $u(t)$.
In the white noise limit, Eq.~\eqref{eq:R} becomes:
\begin{equation} \label{eq:Rwhite}
\frac{\partial^2}{\partial s^2}R(s,r,\mu) +
\upsilon^2\mu\, R(s,r,\mu)  \; = \; -
\upsilon^2\,\delta (s-r),
\end{equation}
with  $R(0,r,\mu) = R(t,r,\mu) = 0$ for any $r, \mu$. $R(s,r,\mu)$
clearly displays a discontinuity on its first derivative for $s=r$.
The solution is found by imposing on the solution of
the homogeneous equation the correct discontinuity conditions on the
first derivative, together with the boundary conditions.
$R(s,r,\mu)$, having the correct boundary conditions in $s=0$ and
$s=t$, is easily determined:
\begin{equation}
R(s,r,\mu)=\left\{
\begin{array}{l}
\displaystyle a \sinh i\upsilon\sqrt{\mu}s\quad\qquad s\leq r\,,
\\ \\ \displaystyle
b \sinh i\upsilon\sqrt{\mu}(s-t)
\quad\qquad s\geq r\,,
\end{array}\right.
\end{equation}
where $a$ and $b$ are two constants determined by the following
relations:
\begin{equation}
\left\{\begin{array}{l}
\displaystyle a \sinh i\upsilon\sqrt{\mu}r= b
\sinh i\upsilon\sqrt{\mu}(r-t)\,,\\ \\ \displaystyle
b\sqrt{\mu} \cosh i\upsilon\sqrt{\mu}(r-t)
-a\sqrt{\mu}\cosh i\upsilon\sqrt{\mu}r =-\frac{2i}{m\upsilon}\,,
\end{array}\right.
\end{equation}
the first equation represents the continuity  of $R(s,r,\mu)$ in $r$
and the second equation encodes the discontinuity of the first
derivative.

\noindent After some algebra on finds the following result:
\begin{equation}
R(s,r,\mu)=\left\{
\begin{array}{c}
\displaystyle \frac{i\upsilon}{\sqrt{\mu}} \frac{\sinh
i\upsilon\sqrt{\mu}(r-t)\,\, \sinh i\upsilon\sqrt{\mu}s}{\sinh
i\upsilon\sqrt{\mu}t}\quad\qquad s\leq r, \\ \\ \displaystyle
\frac{i\upsilon}{\sqrt{\mu}} \frac{\sinh i\upsilon\sqrt{\mu}r
\,\,\sinh i\upsilon\sqrt{\mu}(s-t)}{\sinh i\upsilon\sqrt{\mu}t}
\quad\qquad s\geq r.
\end{array}\right.
\end{equation}
One then has:
\begin{equation}
\int_{-1}^0d\mu\int_0^tdr\,R(r,r,\mu)=\left[\ln\frac{\sqrt{\mu}}{\sinh
i\upsilon\sqrt{\mu} t}\right]_{-1}^0\,;
\end{equation}
the above formula, when evaluated in $\mu=-1$, produces an extra $i$
factor which cancels the one in the argument of $\sinh$.
Substituting this result into Eq.~\eqref{eq:det}, one  finds that:
\begin{equation}\label{eq:uwhite}
u(t) = \frac{\sinh\upsilon t}{\upsilon t}\,.
\end{equation}
Substituting Eq.~\eqref{eq:uwhite} into~\eqref{eq:prop} the
prefactor in front of the Green's function becomes:
\begin{equation}
\sqrt{\frac{\lambda}{\pi\upsilon\sinh\upsilon t}}\,,
\end{equation}
which is the prefactor known in literature
~\cite{Kolokoltsov:98,Grosche:98,Bassi:08}.

\section{Exponential correlation function}
\label{sec:six} The results of  section \ref{sec:three} are
important by themselves. However, they acquire  a further importance
when Eqs.~\eqref{eq:z} and~\eqref{eq:R} can be solved for some
specific, physically meaningful, correlation function. In this
section we treat the case in which the correlation function has an
exponential damping, i.e.
\begin{equation}\label{eq:expcorr}
D(t,s)=\frac{\gamma}{2}e^{-\gamma |t-s|}\,,
\end{equation}
where $\gamma$ is the inverse of the correlation time. This
type of correlation function was first considered in~\cite{Diosi:98}. Since
$D(t,s)$ is time translation invariant, we can use the
expression~\eqref{eq:ansnogh} for $z(s)$. The equation for $f(s)$
reads:
\begin{equation}\label{eq:f}
f''(s)-\frac{i\gamma\omega^2}{2}\int_0^t dr\, e^{-\gamma |s-r|}f(r) \; = \;0\,,
\end{equation}
with $\omega$ defined~\eqref{eq:fwhite}. The equation has to be solved with the boundary conditions $f(0)=1$, $f(t)=0$. The
solution can be found using the following mathematical
procedure~\cite{Polyanin:08}: one differentiates the equation twice,
obtaining:
\begin{equation}
f''''(s)+i\gamma^2\omega^2f(s)+\gamma^2\left(-\frac{i\gamma\omega^2}{2}\int_0^s
dl\, e^{-\gamma (s-l)}f(l)-\frac{i\gamma\omega^2}{2}\int_s^t dl\,
e^{\gamma (s-l)}f(l)\right)=0\,;
\end{equation}
next, using once again Eq.~\eqref{eq:f}, one replaces the third term
with $-\gamma^2f''(s)$. In this way Eq.~\eqref{eq:f} is transformed
into the following fourth-order differential equation:
\begin{equation}\label{eq:f4}
f''''(s)-\gamma^2f''(s)+i\gamma^2\omega^2f(s)=0\,.
\end{equation}
The general solution of this equation, which has the advantage of
not having an integral term, is easily found and can be expressed as
a sum of hyperbolic sines and cosines as follows:
\begin{equation}\label{eq:fgen}
f(s)=f_1 \sinh \upsilon_1 s +f_2\sinh \upsilon_2 s +f_3 \cosh \upsilon_1 s
+f_4 \cosh \upsilon_2 s\,,
\end{equation}
where $f_i$ are constants, to be determined by the boundary
conditions, while $\upsilon_i$ are the two non-symmetric roots of
the bi-quadratic characteristic polynomial associated to
Eq.~\eqref{eq:f4}, i.e.
\begin{equation} \label{eq:gdsfsdasda}
\upsilon_{1,2}=\sqrt{\frac{1}{2}\left(\gamma^2\pm\zeta\right)}\,,
\qquad\qquad\zeta=\sqrt{\gamma^4-4i\gamma^2\omega^2}\,;
\end{equation}
the other two solutions are: $\upsilon_3=-\upsilon_1$ and
$\upsilon_4=-\upsilon_2$. It is easy to verify that in the white
noise limit ($\gamma\rightarrow\infty$): $\upsilon_1\rightarrow
O(\gamma)$ and $\upsilon_2\rightarrow\upsilon$ defined in~\eqref{eq:fwhite}.

Since Eq.~\eqref{eq:f4} is a fourth-order differential equation, to
determine a specific solution we need four boundary conditions.
Two of them are $f(0)=1$ and $f(t)=0$. In order to find the other
two conditions, we proceed as in~\cite{Polyanin:08}. We evaluate
Eq.~\eqref{eq:f} in $0$ and in $t$:
\begin{equation}
\left\{
\begin{array}{l}
\displaystyle f''(0)-\frac{i\gamma\omega^2}{2}\int_0^t dl\,
e^{-\gamma l}f(l)\,=\,0\\
\\
\displaystyle f''(t)-\frac{i\gamma\omega^2}{2}\int_0^t dl\,
e^{-\gamma (t-l)}f(l)\,=\,0\,,
\end{array}\right.
\end{equation}
and we replace $f(l)$ in each expression, using Eq.~\eqref{eq:f4}:
\begin{equation}
\left\{
\begin{array}{l}
\displaystyle f''(0)+\frac{1}{2\gamma}\int_0^t dl\,
e^{-\gamma l}(f''''(l)-\gamma^2f''(l))\,=\,0\\
\\
\displaystyle f''(t)+\frac{1}{2\gamma}\int_0^t dl\,
e^{-\gamma (t-l)}(f''''(l)-\gamma^2f''(l))\,=\,0\,.
\end{array}\right.
\end{equation}
Integrating by parts these two equations, after some algebra one
finds that the complete set of boundary conditions is:
\begin{equation}
\left\{
\begin{array}{cccc}
\displaystyle f(0)=1\,,& & & \displaystyle
f'''(0)=\gamma f''(0)\\ \\ \displaystyle
f(t)=0\,,& & &\displaystyle
f'''(t)=-\gamma f''(t)\,.
\end{array}\right.
\end{equation}

The solution of this system involves a long calculation and leads to
a complicated expression for the coefficients $f_i$
in~\eqref{eq:fgen}. In order to write the solution $f(s)$ in a
compact way, we introduce the following new coefficients:
\begin{eqnarray}
a_k &=& \gamma \upsilon_k^3[\upsilon_k^2+(-1)^{\bar{k}} \zeta]\,,\\
b_k&=&\upsilon_k^2[\upsilon_k^4+(-1)^{\bar{k}}\gamma^2\zeta]\,,\\
c&=&\upsilon_1^3\upsilon_2^3\,,\\
d_k&=&-\gamma \upsilon_k^3\upsilon_{\bar{k}}^2\,,
\end{eqnarray}
where $k=1,2$ and $\bar{k}=2$ if $k=1$, $\bar{k}=1$ if $k=2$. With the help of
these prescriptions, $f(s)$ can be written as follows:
\begin{equation}\label{eq:fsol}
f(s) \; = \; f_t(s) \; = \; \frac{\sum_k\left[r_t^k \sinh \upsilon_k(t-s) +
u_t^k \cosh \upsilon_k(t-s) - u_s^k\right]}{ \sum_{k}\left[2c + r_t^k \, \sinh \upsilon_k t
+ u_t^k \cosh \upsilon_k t\right]}\,,
\end{equation}
where:
\begin{eqnarray}
r_t^k & = & a_{\bar{k}} \cosh \upsilon_{\bar{k}}t
+b_{\bar{k}}\sinh \upsilon_{\bar{k}}t \,,\\
u_t^k & = &d_k\sinh \upsilon_{\bar{k}}t-c \cosh \upsilon_{\bar{k}}t \,.
\end{eqnarray}

Resorting once again to time translation invariance, we know that
$h(s)$ takes the form~\eqref{eq:gfhsd}, when the particular solution
$h^{\text{\tiny P}}(s)$ has the form~\eqref{eq:dfgdsfp}. In order to
find $h^{\text{\tiny P}}(s)$, we have to consider the
non-homogeneous equation:
\begin{equation}\label{eq:ffgsdf}
h''(s)-\frac{i\gamma\omega^2}{2}\int_0^t dr\, e^{-\gamma |s-r|}h(s)
\; = \; -\frac{i\sqrt{\lambda}\hbar}{m} w(s)\,.
\end{equation}
Following the same calculations which brought from Eq.~\eqref{eq:f}
to Eq.~\eqref{eq:f4}, one finds the following fourth-order
non-homogeneous equation for $h(s)$:
\begin{equation}\label{eq:h4}
h''''(s)-\gamma^2h''(s)+i\gamma^2 \omega^2 h(s)=
-\frac{i\sqrt{\lambda}\hbar}{m}\left(w''(s)-\gamma^2w(s)\right)\,.
\end{equation}
A particular solution of this equation, having the
form~\eqref{eq:dfgdsfp}, is~\cite{Hartman:02}:
\begin{equation}\label{eq:h}
h^{\text{\tiny P}}(s) = -\frac{i\sqrt{\lambda}\hbar}{m}\int_0^s
\bar{f}_s(l)\left(w''(l)-\gamma^2w(l)\right)dl\,,
\end{equation}
where $\bar{f}_s(l)$ is the solution of Eq.~\eqref{eq:f4}, thus
having the general form~\eqref{eq:fgen}, with boundary conditions:
$f_s(s)=f_s'(s)=f_s''(s)=0$, $f_s'''(s)=1$. After some simple
calculations, one finds the following expression:
\begin{equation}
\displaystyle \bar{f}_s(l)\,=\,-\frac{1}{\zeta}\sum_k(-1)^k\frac{\sinh
\upsilon_k(s-l)}{\upsilon_k}\,.
\end{equation}
We now have all functions which are necessary to compute the Green's
function. This takes the form~\eqref{eq:propts}, where the
coefficients $\mathcal{A}_t\,$--$\,\mathcal{E}_t$ are defined in
Eqs.~\eqref{eq:gvxshnj}--\eqref{eq:gvxshnj2} in terms of the
following functions:
\begin{eqnarray} \label{eq:luojfgcxgf}
f_t'(0)&=&-\frac{\sum_k\upsilon_k\big(r_t^k \cosh \upsilon_kt  +
u_t^k \sinh \upsilon_kt + d_{\bar{k}}\big)}{ \sum_k\left[2c + r_t^k \, \sinh \upsilon_k t
+ u_t^k \cosh \upsilon_k t\right]}\,,\\
f_t'(t)&=&-\frac{\sum_k\upsilon_k\big(r_t^k + d_{\bar{k}}\cosh \upsilon_kt -c \sinh \upsilon_kt\big)}{\sum_k\left[ 2c + r_t^k \, \sinh \upsilon_k t
+ u_t^k \cosh \upsilon_k t\right]}\,,
\end{eqnarray}
and:
\begin{eqnarray}
h^{\text{\tiny P}'}(0)&=&0\\
h^{\text{\tiny P}'}(t)&=&\frac{i\sqrt{\lambda}\hbar}{m\zeta}\int_0^t \sum_k(-1)^k\cosh\upsilon_k (t-l)\left(w''(l)-\gamma^2w(l)\right)\,.
\end{eqnarray}
In terms of these functions, the entire dynamics of the free
particle can be analyzed.

Eq.~\eqref{eq:R} for $R(s,r,\mu)$ can be solved similar to Eq.~\eqref{eq:z}. One then finds
the explicit expression of the prefactor of the Green's
function~\eqref{eq:propts}. This goes outside the scope of the
present paper, so we omit the calculation.

\section{Density matrix evolution and imaginary noise trick}
\label{sec:seven}

One of the major difficulties connected to Eq.~\eqref{eq:main}
arises when one wants to compute average values of the form ${\mathbb E}_{\mathbb P} [ \langle \psi_{t} | O | \psi_{t} \rangle ]
\equiv {\mathbb E}_{\mathbb Q} [ \langle \phi_{t} | O | \phi_{t}
\rangle ]$, where $\phi_{t}$ solves Eq.~\eqref{eq:main} and $\psi_t$
is the corresponding normalized solution; $O$ is a generic
self-adjoint operator. Averages of this kind are particularly
important, as they represent physical quantities, directly connected
to experimental outcomes. The difficulty in computing such averages
lies both in the difficulty in solving Eq.~\eqref{eq:main} and in
the fact that $\phi_t$ depends non-trivially on the noise $w(t)$.

In the white noise-case, a very helpful trick, known as the {\it
imaginary noise trick}~\cite{Ghirardi:90,Adler:07} allows to simplify considerably the
problem. Let us consider the following class of stochastic
differential equations:
\begin{equation} \label{eq:cpe}
d\psi_{t}^{\xi} = \left[ - \frac{i}{\hbar} H dt + \sqrt{\gamma}
\sum_{n = 1}^{N} (\xi q_{n} - \xi_{\text{\tiny R}} \langle q_{n}
\rangle_{t}) d W_{n,\,t} - \frac{\gamma}{2} \sum_{n=1}^{N} (|\xi|^2
q_{n}^2 - 2\xi\xi_{\text{\tiny R}} q_{n} \langle q_{n} \rangle_{t} +
\xi_{\text{\tiny R}}^2 \langle q_{n} \rangle_{t}^2) dt \right]
\psi_{t}^{\xi},
\end{equation}
where $\xi = \xi_{\text{\tiny R}} + i \xi_{\text{\tiny I}} $ is a
constant complex factor. By using standard It\^o calculus one can
show that ${\mathbb E}_{\mathbb P} [ \langle \psi_{t}^{\xi} | O |
\psi_{t}^{\xi} \rangle ]$ depends only on $|\xi|^2$: there is, so to
speak, an invariance under phase change in the coupling constant.
When $\xi = 1$ one recovers Eq.~\eqref{eq:nlemp}, while when $\xi =
i$, one obtains a standard Schr\"odinger equation with a stochastic
potential, generating a unitary (thus non-collapsing) evolution,
without non-linear terms; this equation is much simpler to analyze
than Eq.~\eqref{eq:nlemp}, nevertheless, at the statistical level,
it gives the same results as Eq.~\eqref{eq:nlemp} does.

In~\cite{Adler:07} it has been shown that the imaginary noise trick holds also
in the non-white noise case, at least to second order in the
perturbation expansion with respect to the parameter
$\sqrt{\lambda}$. Aim of this section is to prove that this is an
exact property which holds to all orders.

Let us consider the following class of equations:
\begin{equation} \label{eq:evxi}
\frac{d}{dt}\, \phi_t^{\xi}(x) \; = \; \left[ - \frac{i}{\hbar}\, H
\, + \, \xi \sqrt{\lambda} q w(t) \, - \, 2\xi_{\text{\tiny R}}
\sqrt{\lambda} q \int_0^t ds D(t,s) \frac{\delta}{\delta w(s)}
\right] \phi_t^{\xi}(x)\,;
\end{equation}
obviously, when $\xi=1$ one recovers Eq.~\eqref{eq:freepart} for a
free particle. Since ${\mathbb E}_{\mathbb P} [ \langle \psi_{t} | O
| \psi_{t} \rangle ] \equiv {\mathbb E}_{\mathbb Q} [ \langle
\phi_{t} | O | \phi_{t} \rangle ] \equiv \text{Tr}[ O \rho_t]$ where
the density matrix $\rho_t$ is defined as:  $\rho_t := {\mathbb
E}_{\mathbb P} [ | \psi_t \rangle \langle \psi_t |] \equiv {\mathbb
E}_{\mathbb Q} [ | \phi_t \rangle \langle \phi_t |]$, in order to
prove the required property it is sufficient to show that $\rho_t$
depends only on $|\xi|^2$. The propagator associated to
Eq.~\eqref{eq:evxi} reads
\begin{equation} \label{eq:fgfytf}
G_{\xi}(x,t;x_0,0) \; = \; \int^{q(t)=x}_{q(0)=x_0} \mathcal{D}[q]
e^{\mathcal{S}_{\xi}[q]} \,,
\end{equation}
where
\begin{equation} \label{eq:bvdxfg}
\mathcal{S}_{\xi}[q] \; = \; \int^t_0 ds \left[\frac{i}{\hbar}
\mathcal{S}_{0}[q] + \xi\sqrt{\lambda} q(s) w(s) - \xi_{\text{\tiny
R}}\xi\lambda q(s) \int_0^t dr\, q(r) D(s,r)\right]\,,
\end{equation}
and $\mathcal{S}_{0}[q]$ is the standard action associated to the quantum Hamiltonian
$H$. The propagator $J(x,x',t;x_0,x'_0,0)$  associated to
$\rho_t(x,x') := \langle x | \rho_t | x' \rangle$ can then be
expressed as follows~\cite{Feynman:65}:
\begin{equation} \label{eq:fsddfd}
J(x,x',t;x_0,x'_0,0)={\mathbb E}_{\mathbb Q}
\left[G_{\xi}(x,t;x_0,0)G_{\xi}^*(x',t;x'_0,0)\right]\,,
\end{equation}
where ${}^{*}$ denotes complex conjugation. Substituting the
definition~\eqref{eq:fgfytf} for the wave function's propagator
into~\eqref{eq:fsddfd}, and exchanging the path-integration with the
stochastic average, the stochastic terms average as follows:
\begin{eqnarray}
\lefteqn{{\mathbb E}_{\mathbb Q} \left[\exp\left[\int_0^t
ds\left(\xi\sqrt{\lambda} q(s)+\xi^{*}\sqrt{\lambda}
q'(s)\right)w(s)\right]\right]=} \nonumber \\
& = & \exp\left[\frac{1}{2}\int_0^tds\int_0^tdr D(s,r)
\left(\xi^2\lambda q(s)q(r)+\xi^{* 2}\lambda q'(s)q'(r)+
|\xi|^2\lambda q(s)q'(r)+|\xi|^2\lambda q(r)q'(s)\right)\right]\,.
\nonumber \\ & &
\end{eqnarray}
The propagator then becomes:
\begin{eqnarray}
J(x,x',t;x_0,x'_0,0)&=&\int^{q(t)=x}_{q(0)=x_0} \mathcal{D}[q]
\int^{q'(t)=x'}_{q'(0)=x'_0} \mathcal{D}[q']\exp\left[\int_0^tds
\bigg(\frac{i}{\hbar} \mathcal{S}_{0}[q]-\frac{i}{\hbar} \mathcal{S}_{0}[q']\right.\nonumber\\
&&\qquad\qquad\quad\left.\left.-\frac{\lambda}{2}|\xi|^2\int_0^tdr
D(s,r)
\left(q(s)-q'(s)\right)\left(q(r)-q'(r)\right)\right)\right]\,,
\end{eqnarray}
which depends only on $|\xi|^2$. We can conclude that the evolution
for the density matrix is independent from a phase change in the
coupling with the noise.

\subsection{Average values of physical quantities}

As anticipated, this phase change invariance provide a very handy
tool to compute average values, such as the mean position and mean
momentum. In place of Eq.~\eqref{eq:freepart}, let us consider the
equation:
\begin{equation} \label{eq:unit}
\frac{d}{dt}\, \phi_t(x) \; = \; \left[ - \frac{i}{\hbar}\,
\frac{p^2}{2 m} \, + \, i \sqrt{\lambda} q w(t)\right] \phi_t(x)\,,
\end{equation}
which belongs to the class~\eqref{eq:evxi}, with $\xi=i$. This is a
linear, unitary, norm-preserving standard Schr\"odinger equation for
a free particle under the influence of a stochastic potential. The
evolution of the stochastic average of the mean value of an
observable $O$ is given by the following equation:
\begin{equation}\label{eq:opunit}
\frac{d}{dt}{\mathbb E}_{\mathbb Q} \left[\langle
O\rangle_t\right]=\frac{i }{2m\hbar}{\mathbb E}_{\mathbb Q} [\langle
[p^2,O] \rangle_t] - i\sqrt{\lambda} {\mathbb E}_{\mathbb Q}
[w(t)\langle\left[q,O\right] \rangle_t]\,,
\end{equation}
where, as usual, $\langle \cdot \rangle_t = \langle\phi_t
|\cdot|\phi_t\rangle$.

The mean in position and momentum can now be easily computed.
Remembering that $\langle \phi_t | \phi_t \rangle = 1$ and that the
mean of  $w(t)$ is 0, one finds:
\begin{eqnarray}
\frac{d}{dt}{\mathbb E}_{\mathbb Q} \left[\langle q\rangle_t\right]
& = & \frac{1}{m}{\mathbb E}_{\mathbb Q} \left[\langle
p\rangle_t\right]\,, \label{eq:yhdfgdfgq}\\
\frac{d}{dt}{\mathbb E}_{\mathbb Q} \left[\langle p\rangle_t\right]
& = & 0\,; \label{eq:yhdfgdfgp}
\end{eqnarray}
as we see, we recover Newton's equations for a free particle. In
particular, also in the non-Markovian case, like in the white-noise
case, the momentum of an isolated system is conserved, in the
average. In sec.~\ref{sec:eight}, where we analyze the time evolution of
Gaussian states, we will see that the fluctuations around the
average are inversely proportional to the mass of the system. These
two facts, together with the fact that the collapse scales with
the size of the system, lead to the following result: {\it the wave
function of an isolated macro-object behaves, for all practical
purposes, like a particle moving deterministically in space
according to Newton's laws.}

Because of the many experimental implications~\cite{Adler3:07},  it
is important to check how the mean free energy $H_0 = p^2/2m$
evolves in time. This is most easily computed by shifting to the
Heisenberg picture, in which case one finds~\cite{Adler:07}:
\begin{equation} \label{eq:dcxgu}
\frac{d}{dt}{\mathbb E}_{\mathbb Q} \left[\langle H_0
\rangle_t\right] =\frac{\lambda\hbar^2}{m}\int_0^t ds D(t,s)\,;
\end{equation}
this is the expected generalization of the well-known white noise
formula for the mean energy increase in collapse
models~\cite{Ghirardi:86,Ghirardi2:90,Bassi2:05}. For a physically
reasonable correlation function such as the exponential one of
Eq.~\eqref{eq:expcorr}, one has:
\begin{equation}
{\mathbb E}_{\mathbb Q} \left[\langle H_0 \rangle_t\right] \; =
\;\langle H_0 \rangle_0+\frac{\lambda\hbar^2}{2m}\left(t+
\frac{e^{-\gamma t}-1}{\gamma}\right)\,:
\end{equation}
as we see, also in this case the mean energy increases linearly in
time, without reaching a steady value. More generally, let us assume
time translation invariance, and let us consider the spectral
decomposition of a generic correlator $D(|t-s|)$:
\begin{equation}
D(|t-s|) \; = \; \int_0^{\infty} d\omega \tilde{D}(\omega) \cos
\omega (t-s);
\end{equation}
we have:
\begin{equation}
\int_0^t ds D(t,s) \; = \; \int_0^{\infty} \frac{du}{u}
\tilde{D}(u/t) \sin u \;\; \xrightarrow[t \,\rightarrow \,
+\infty]{} \;\; \frac{\pi}{2} \tilde{D}(0).
\end{equation}
The above formula shows that the energy production is nonzero also
at large times, unless the correlator has a cutoff at low
frequencies~\cite{Adler:07}.

An important lesson to learn from the above analysis is that
non-Markovian terms do not introduce thermalization effects in the
evolution of a quantum system; such effects can be introduced only
by modifying the form of the operator coupled to the noise, as first
discussed in~\cite{Bassi3:05}. Nevertheless, non-Markovian terms are
extremely important, as they affect the time evolution of physical
quantities, thus the predictions of collapse models, at small times.
A significative example has been first provided in~\cite{Adler2:07}.

\section{Evolution of a gaussian wave function}
\label{sec:eight}

In this section we study the time evolution of Gaussian wave
functions, whose form is clearly preserved by the Green's
function~\eqref{eq:propexp}; for simplicity, we will assume time
translation invariance, so that~\eqref{eq:propexp} reduces
to~\eqref{eq:propts}. A generic Gaussian state of the form:
\begin{equation}
\phi_0(x_0)=\exp[-\alpha_0x_0^2+\beta_0x_0+\gamma_0]\,.
\end{equation}
evolves in time to the following state:
\begin{equation} \label{eq:gfsdss}
\phi_t(x)=\exp[-\alpha_tx^2+\beta_tx+\gamma_t]\,,
\end{equation}
where:
\begin{eqnarray} \label{eq:fsadfsaf}
\alpha_t & = &
\mathcal{A}_t-\frac{\mathcal{B}^2_t}{4(\alpha_0+\mathcal{A}_t)},\\
\beta_t & = &
-\frac{\mathcal{C}_t+\beta_0}{2(\alpha_0+\mathcal{A}_t)}\mathcal{B}_t+\mathcal{D}_t,\\\gamma_t & = &
\gamma_0+\mathcal{E}_t+\frac{(\mathcal{C}_t+\beta_0)^2}{4(\alpha_0+\mathcal{A}_t)}\,,
\end{eqnarray}
and the functions $\mathcal{A}_t$ -- $\mathcal{E}_t$ have been
defined in Eqs.~\eqref{eq:gvxshnj}--\eqref{eq:gvxshnj2}.  We can
draw some important conclusions about the time evolution of Gaussian
wave functions:

\begin{itemize}
\item
Since neither $\mathcal{A}_t$ nor $\mathcal{B}_t$ depend on the
noise $w(t)$, the time evolution of the spread of the wave function
is deterministic, as in the white noise case~\cite{Bassi2:05}. An
initially spread-out Gaussian wave function shrinks in space,
reaching an asymptotic final spread. In the case of an exponential
correlation function, we can provide the explicit expression for the asymptotic value for the
spread (see next subsection).

\item By making the substitution:
\begin{equation} \label{eq:hdssdd}
x \; \rightarrow \; y \; := \; \sqrt{\frac{m}{m_0}}\, x\,,
\end{equation}
one can easily prove that Eq.~\eqref{eq:freepart}, thus the
propagator~\eqref{eq:propts}, does not depend on the mass $m$ of the
particle. Thus, one way to see the effect of the mass on the global
dynamics is to take the evolution of the wave function for the
reference mass $m_0$, and then ``shrink'' the space coordinates by a
factor $\sqrt{m_0/m}$. This leads to the {\it amplification
mechanism}, which is the characteristic feature of collapse models:
the bigger the system, the faster the wave function shrinks in
space.

\item
The center $\langle q \rangle_t = \beta_t^{\text{\tiny R}}/2
\alpha_t^{\text{\tiny R}}$ of the Gaussian wave
function~\eqref{eq:gfsdss} evolves {\it randomly} in space, as
expected. Its average value has already been computed
in~\eqref{eq:yhdfgdfgq}, and evolves classically. Moreover, due to
the independence of the dynamics from $m$ under the
substitution~\eqref{eq:hdssdd}, one can conclude that the
fluctuations of $\langle q \rangle_t$ around the classical motion go
like $m^{-1/2}$ since the following equality for the variance
$\mathbb{V}_q := \sqrt{\mathbb{E}_{\mathbb{Q}}\left[\langle
q\rangle_t- \mathbb{E}_{\mathbb{Q}}[\langle q\rangle_t]\right]^2}$
follows immediately from~\eqref{eq:hdssdd}:
\begin{equation}\label{eq:var}
\mathbb{V}_q^m\; = \; \sqrt{\frac{m_0}{m}}\; \mathbb{V}_q^{m_0}\,,
\end{equation}
where the apex $m$ indicates with respect to which mass the variance is computed. Eq.~\eqref{eq:var}
in agreement with the white noise case. This means
that the bigger the system, the smaller the fluctuations: in the
microscopic case, in this way, one recovers classical determinism for all pratical pourposes.

\item
The mean momentum $\langle p \rangle_t = \hbar[\beta_t^{\text{\tiny
I}}-(\alpha_t^{\text{\tiny I}}/\alpha_t^{\text{\tiny
R}})\beta_t^{\text{\tiny R}}]$ also evolves randomly in space. Its
average value is constant in time (see Eq.~\eqref{eq:yhdfgdfgp}), as
expected from a free particle, while the fluctuations around the
average increase like $m^{1/2}$. If however we consider the
fluctuations of the mean velocity, we have that they decrease like
$m^{-1/2}$. Thus also in this respect one recovers classical
determinism at the macroscopic level.
\end{itemize}
The above remarks show that the non-Markovian QMUPL model shares all
the important features of the white noise model. More quantitative
details can be given by taking a specific expression for the
correlation function $D(t,s)$, as we will do in the next subsection.

\subsection{Exponential correlation function}
\label{sec:nine}

\begin{figure}[t]
\begin{center}
\includegraphics[width=0.5\textwidth]{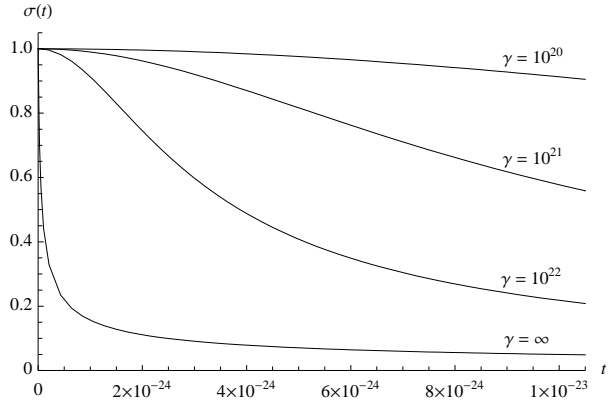}
\caption{Linear/linear graph showing the time evolution of
$\sigma(t)$ for a 1 Kg particle, for different values of $\gamma$,
for the same initial condition $\sigma(0) = 1$ and with $\lambda_0 = 5.00
\times 10^{-3}$ m$^{-2}$ sec$^{-1}$. The curve with $\gamma = \infty$
corresponds to the white noise case. Time is measured in seconds,
distances in meters.} \label{fig:1}
\end{center}
\end{figure}

\begin{figure}
\begin{center}
\includegraphics[width=0.5\textwidth]{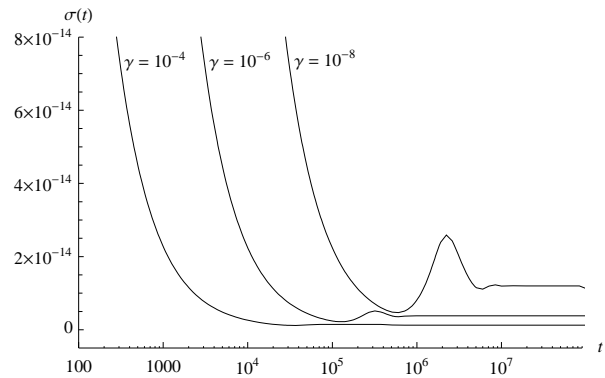}
\caption{Log/linear graph, showing the large time behavior of
$\sigma(t)$ for a 1 Kg particle, for different values of $\gamma$,
for the same initial condition $\sigma(0) = 1$ and with $\lambda_0 =5.00\times
10^{-3}$ m$^{-2}$ sec$^{-1}$. The white noise case would appear in
the graph as a straight line with value 1.27 $\times 10^{-15}$ m.
Time is measured in seconds, distances in meters.} \label{fig:2}
\end{center}
\end{figure}

We study in some more detail the time evolution of the Gaussian
solution~\eqref{eq:gfsdss}, when the correlation function is
exponential as in~\eqref{eq:expcorr}. We focus our attention on the
spread in position $\sigma(t)$ which is given by the inverse of
twice the square root of the real part of $\alpha_t$, whose analytic
expression is given explicitly by
Eqs.~\eqref{eq:fsadfsaf},~\eqref{eq:gvxshnj}
and~\eqref{eq:luojfgcxgf}. Figs.~\ref{fig:1} and~\ref{fig:2} show
some typical behavior of $\sigma(t)$ for small times
(Fig.~\ref{fig:1}) and large times (Fig.~\ref{fig:2}), for different
values of $\gamma$ and for $m = 1$ Kg. As we see and as we expect,
the larger the value of $\gamma$, the stronger the noise and the
faster the collapse in space.

\begin{figure}
\begin{center}
\includegraphics[width=0.5\textwidth]{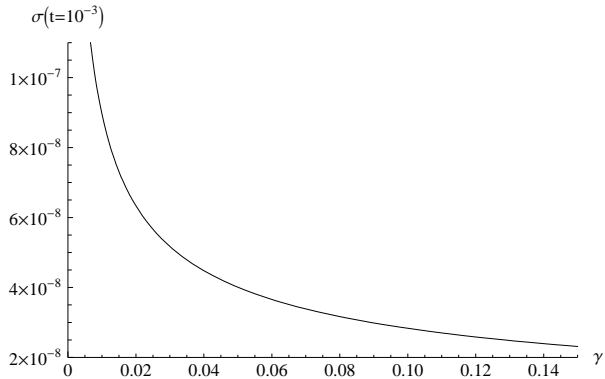}
\caption{Linear/linear graph, showing to which value an initial
spread $\sigma(0) = 1$ m decreases after $10^{-3}$ sec, as a
function of $\gamma$. The mass of the particle has been set equal to
$m =1.01\times 10^{-3}$ Kg, corresponding to the total mass of a system
containing an Avogadro's number of nucleons. The coupling constant
$\lambda_0$ has been given the GRW-value $5.00 \times10^{-3}$ m$^{-2}$
sec$^{-1}$. Larger values of $\gamma$ imply a faster collapse in
space. Time is measured in seconds, distances in meters.} \label{fig:3}
\end{center}
\end{figure}

Fig.~\ref{fig:3} shows how the spread of a Gaussian wave
function associated to a particle having a mass $m = 1.01\times 10^{-3}$ Kg
(which is the total mass of a system of $6.02\times 10^{23}$ nucleons), having
initial spread $\sigma(0) = 1$ m, decreases after $10^{-3}$ sec, as
a function of $\gamma$. The value of the coupling constant
$\lambda_0$ is the GRW value~\eqref{eq:cvxvxc}.
GRW have chosen the value of the coupling constant in order to
ensure that the wave function of a system of at least an Avogadro's
number of particle collapses, within a time interval of $10^{-3}$
sec, below $10^{-7}$ m. Fig.~\ref{fig:3} shows that also for
relatively small values of $\gamma$, the non-Markovian collapse
models preserves this same feature of the white-noise model.

\begin{figure}
\begin{center}
\includegraphics[width=0.5\textwidth]{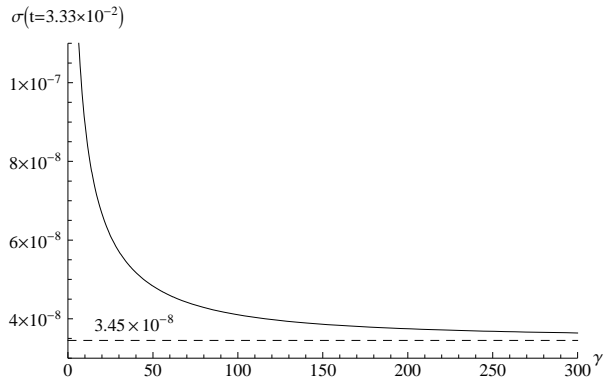}
\caption{Linear/linear graph, showing to which value an initial
spread $\sigma(0) = 1$ m decreases after $3.33\times 10^{-2}$ sec,
as a function of $\gamma$. The mass of the particle has been set
equal to $m = 1.06\times 10^{-18}$ Kg, which is the mass of a system
of $n^2N$ ($n=5640$, $N=20$) nucleons, with $n$ and $N$ taken from
Eq.~(8) of~\cite{Adler3:07}. The coupling constant $\lambda_0$ has
been given Adler's middle value $1.12\times10^{6}$ m$^{-2}$
sec$^{-1}$. The dashed line corresponds to the white noise case.
Time is measured in seconds, distances in meters.} \label{fig:4}
\end{center}
\end{figure}

Fig.~\ref{fig:4} displays the same graph as that of
Fig.~\ref{fig:3}, with the typical values chosen by Adler.
In~\cite{Adler3:07} Adler sets the value of the CSL-coupling
constant $\gamma_{\text{\tiny{CSL}}}$ equal to $2\times
10^{-21\pm2}$ cm$^3$ sec$^{-1}$ by noting that in the process of
latent image formation, which has a characteristic time of about
$3.33 \times 10^{-2\pm2}$ sec, a number of atoms approximately equal
to 20 is involved. He then assumes that the collapse process ensures
that the reduction occurs already at this stage, from which the
value $\gamma_{\text{\tiny{CSL}}} = 2\times 10^{-21\pm2}$ cm$^3$
sec$^{-1}$ comes. The graph shows that even for relatively small
values of $\gamma$ (Adler and Ramazanoglu~\cite{Adler2:07} have
shown that choosing $\gamma \sim 10^{15}$ sec$^{-1}$ already changes
significantly the predictions of non-Markovian models with respect
to white-noise models), an initially spread-out wave function
shrinks rapidly below $10^{-7}$ m, which corresponds to a
well-localized wave packet.

{}For an exponential correlation function, we can write down the
analytic expression of the asymptotic value of $\alpha_t$, which is:
\begin{equation} \label{eq:iiidstf}
\alpha_{\infty}=\lim_{t\rightarrow\infty}\alpha_t = -\frac{i
m}{2\hbar}(\upsilon_1+\upsilon_2-\gamma)\,,
\end{equation}
with $\upsilon_{1,2}$ defined in~\eqref{eq:gdsfsdasda}. From this,
the asymptotic spread in position can be easily computed. In
particular, if we take the white noise limit
$\gamma\rightarrow\infty$, one obtains the finite value:
\begin{equation}
\alpha_{\infty}^{\text{\tiny WN}}=\sqrt{\frac{\lambda m}{2 i
\hbar}}\,,
\end{equation}
which matches with the value known in the
literature~\cite{Bassi2:05}.

\section{Conclusions}
\label{sec:ten}

We have thoroughly investigated the dynamics of a free quantum
particle as described by the non-Markovian QMUPL model of
spontaneous wave function collapse. We have provided an explicit
formula for the Green's function; we have shown that it reproduces
the well-known white noise case, and have analyzed the physically
interesting case of an exponential correlation function. We have
proven that the non-Markovian model shares all the most important
features of the corresponding white-noise model; we have described
in particular the evolution of Gaussian wave functions.

There are of course several other important issues which need to be
investigated. In particular:
\begin{itemize}
\item
It is important to set on a rigorous mathematical ground the
change of measure defined in~\eqref{eq:vvbp}, and to derive the
analogous of Girsanov's theorem which holds for the white-noise
case. A sketch of the proof can be found in
Appendix B of Ref.~\cite{Diosi:98}.

\item
It is also important to set a bound on the spread of the general
solution as a function of time, in order to see how it decays in
time. Such a formula would be crucial for setting a lower bound on
the value of $\lambda$ by guaranteeing that the model collapses
sufficiently big systems in sufficiently short time.

\item
In the case of an exponential correlation function, it would be
interesting to prove if {\it any} initial state collapses
asymptotically to a Gaussian state whose spread in position and
momentum is fixed and given by~\eqref{eq:iiidstf}. In this paper we
have proved that this is true only for the special case of initially
Gaussian states. A similar general theorem has been recently proved
for the white-noise case~\cite{Bassi:08}.
\end{itemize}
These problems will be the subject of future investigation.

\section*{Acknowledgments}

The authors wish to thank S.L. Adler, D. D\"urr and G.C.
Ghirardi for many useful conversations. They also wish to thank P.
Pearle for showing them that the Green's function can be equally
well computed by using the standard operator formalism in place of
the path integral formalism. They finally wish to thank A. Fonda for
his guidance through the proof of the existence and uniqueness
theorem of Appendix~\ref{sec:app}.

\appendix
\section{An existence and uniqueness theorem}
\label{sec:app}

In this appendix we prove that equation~\eqref{eq:z}:
\begin{equation}\label{eq:zapp}
\frac{im}{2\hbar}\, z''(s)+\lambda\int_0^t dr\, D(s,r)z(r) \; = \;
\frac{\sqrt{\lambda}}{2}w(s),
\end{equation}
with boundary conditions $z(0)=x_0$, $z(t)=x$, admits a unique
solution. The same theorem applies also to Eq.~\eqref{eq:R}  for
$R(s,r,\mu)$. In order to simplify the proof, it is convenient to
set both boundary conditions to zero. This can be done without loss
of generality, as follows. Let us define the new function:
\begin{equation}\label{eq:zbardef}
\bar{z}(s)= z(s) - y(s)\,,\qquad y(s) := \frac{x-x_0}{t}\,s + x_0\,;
\end{equation}
obviously, $\bar{z}(0)=\bar{z}(t)=0$. Moreover, $\bar{z}(s)$ solves
the following equation:
\begin{equation}\label{eq:zbar}
\frac{im}{2\hbar}\, \bar{z}''(s)+\lambda\int_0^t dr\,
D(s,r)\bar{z}(r) \; = \; \frac{\sqrt{\lambda}}{2}\bar{w}(s)\,,
\end{equation}
where
\begin{equation}
\bar{w}(s)=w(s)-2\sqrt{\lambda}\int_0^t dr\, D(s,r)y(r)\,.
\end{equation}
It is then sufficient to prove the following theorem. \vskip 0.2cm

\noindent \textsc{Theorem.} Let $D(t,s)$ be a real continuous
function on $[0,t]\times [0,t] $, symmetric in its two variables;
let $f(s)$ belong to $C[0,t]$. Then the integro-differential
equation
\begin{equation}\label{eq:th1}
i\, z''(s)+\int_0^t dr\,D(s,r)z(r) \; = \; f(s)\,,
\end{equation}
with boundary conditions $z(0)=z(t)=0$, admits a unique solution
$z(s)\in C^2[0,t]$.\vskip 0.2cm

\noindent \textsc{Proof.} Let $\mathcal{D}(D)\subset C^2[0,t]$ the
set of functions $z(s)\in C^2[0,t]$ such that $z(0)=z(t)=0$. Define
the following operators:
\begin{eqnarray}
D&:&\mathcal{D}(D)\rightarrow C[0,t]\,,\qquad\qquad\qquad\; D[z(s)]\; = \; i z''(s)\,,\\
I&:&C[0,t]\rightarrow C[0,t]\,,\qquad\qquad\qquad I[z(s)]\; = \;
\int_0^t dr\, D(s,r)z(r)\,;
\end{eqnarray}
with these definitions Eq.~\eqref{eq:th1} becomes
\begin{equation}\label{eq:th}
D[z(s)]+I[z(s)]=f(s)\,.
\end{equation}
The integral operator $I$ is compact (see, e.g., Theorem 8.7-5, page
454 of~\cite{Kreyszig:89}), while $D$ is invertible and its inverse
reads:
\begin{eqnarray}
D^{-1}&:&C[0,t]\rightarrow \mathcal{D}(D)\,,\\
D^{-1}[g(s)]&=&
-i\left(\int_0^sdu\,\int_0^udv\,g(v)-\frac{s}{t}\int_0^tdu\,\int_0^udv\,g(v)\right)\,.
\end{eqnarray}
Since it has an integral form, also $D^{-1}$ is compact. We can
write Eq.~\eqref{eq:th} as follows:
\begin{equation}\label{eq:th2}
(\text{Id}-D^{-1}I)[z(s)]=\tilde{f}(s)\,,\qquad z(0)=z(t)=0\,,
\end{equation}
where $\text{Id}$ is the identity operator and
$\tilde{f}(s)=D^{-1}f(s)\in C^2[0,t]$.

Our problem reduces to showing existence and uniqueness of solutions
for Eq.~\eqref{eq:th2}. Since the product of two compact operators
is compact, then also $D^{-1}I$ is compact. We now use Fredholm's
Alternative Theorem (Theorem 8.7-2, page 452 of~\cite{Kreyszig:89}),
according to which in order to prove the theorem it suffices to show
that the homogeneous equation associated to Eq.~\eqref{eq:th2}, i.e.
\begin{equation}\label{eq:thhom}
(\text{Id}-D^{-1}I)[z(s)]=0\,,\qquad z(0)=z(t)=0\,,
\end{equation}
admits only the trivial solution $z(s)=0$.

To prove this, we separate the real and imaginary part of $z(s)$
which we respectively denote with the superscripts
$\text{\tiny{R,I}}$, obtaining
\begin{eqnarray}
(z^{\text{\tiny{R}}})''(s)+\int_0^t dr\, D(s,r)z^{\text{\tiny{I}}}(r)&= &0\,,\\
(z^{\text{\tiny{I}}})''(s)-\int_0^t
dr\, D(s,r)z^{\text{\tiny{R}}}(r) & = &0\,,
\end{eqnarray}
with boundary conditions
$z^{\text{\tiny{R}}}(0)=z^{\text{\tiny{R}}}(t)=z^{\text{\tiny{I}}}(0)=z^{\text{\tiny{I}}}(t)=0$.
Multiplying these two equations respectively by
$z^{\text{\tiny{R}}}(s)$ and $z^{\text{\tiny{I}}}(s)$,
and integrating by parts one finds:
\begin{eqnarray}
\int_0^tds\left[(z^{\text{\tiny{R}}})'(s)\right]^2-\int_0^t ds\,
z^{\text{\tiny{R}}}(s)\int_0^t dr\, D(s,r)z^{\text{\tiny{I}}}(r)&= &0\,,\\
\int_0^t ds
\left[(z^{\text{\tiny{I}}})'(s)\right]^2+\int_0^t
ds\,z^{\text{\tiny{I}}}(s)\int_0^t dr\,
D(s,r)z^{\text{\tiny{R}}}(r) & = &0\,.
\end{eqnarray}
Exploiting the symmetry of $D(s,r)$ in its variables, we can
sum the two equations, obtaining:
\begin{equation}
\int_0^tds
\left[((z^{\text{\tiny{R}}})'(s))^2+((z^{\text{\tiny{I}}})'(s))^2\right]=0\,;
\end{equation}
this implies that $z^{\text{\tiny{R}}}(s)$ and
$z^{\text{\tiny{I}}}(s)$ are constants and, applying the
boundary conditions, these constants are equal to zero.
 %(Here lies the importance of the transformation from $z(s)$ to $\bar{z}(s)$).
We can then state that Eq.~\eqref{eq:thhom} admits only the trivial
solution $z(s)=0$. This proves existence and uniqueness of solutions
for that Eq.~\eqref{eq:th2}, and thus for Eq.~\eqref{eq:z}.

\def\polhk#1{\setbox0=\hbox{#1}{\ooalign{\hidewidth
  \lower1.5ex\hbox{`}\hidewidth\crcr\unhbox0}}} \def\cprime{$'$}


\begin{thebibliography}{10}
\providecommand{\url}[1]{\texttt{#1}}
\providecommand{\urlprefix}{URL }
\providecommand{\eprint}[2][]{\url{#2}}

\bibitem{Diosi:89}
L.~Di\'osi, \emph{Models for universal reduction of macroscopic quantum
  fluctuations}, Phys. Rev. A \textbf{40}, 1165 (1989).

\bibitem{Diosi:90}
L.~Di{\'o}si, \emph{Relativistic theory for continuous measurement of quantum
  fields}, Phys. Rev. A \textbf{42}, 5086 (1990).

\bibitem{Belavkin:89}
V.~P. Belavkin and P.~Staszewski, \emph{A quantum particle undergoing
  continuous observation}, Phys. Lett. A \textbf{140}, 359 (1989).

\bibitem{Belavkin:92}
V.~P. Belavkin and P.~Staszewski, \emph{Nondemolition observation of a free
  quantum particle}, Phys. Rev. A \textbf{45}, 1347 (1992).

\bibitem{Chruscinski:92}
D.~Chru{\'s}ci{\'n}ski and P.~Staszewski, \emph{On the asymptotic solutions of
  {B}elavkin's stochastic wave equation}, Phys. Scripta \textbf{45}, 193
  (1992).

\bibitem{Gatarek:91}
D.~G{\polhk{a}}tarek and N.~Gisin, \emph{Continuous quantum jumps and
  infinite-dimensional stochastic equations}, J. Math. Phys. \textbf{32}, 2152
  (1991).

\bibitem{Halliwell:95}
J.~Halliwell and A.~Zoupas, \emph{Quantum state diffusion, density matrix
  diagonalization, and decoherent histories: a model}, Phys. Rev. D \textbf{52}, 7294 (1995).

\bibitem{Holevo:96}
A.~S. Holevo, \emph{On dissipative stochastic equations in a {H}ilbert space},
  Probab. Theory Relat. Fields \textbf{104}, 483 (1996).

\bibitem{Bassi2:05}
A.~Bassi, \emph{Collapse models: analysis of the free particle dynamics}, J.
  Phys. A \textbf{38}, 3173 (2005).

\bibitem{Bassi:08}
A.~Bassi, D.~D\"urr and M.~Kolb, \emph{On the long time behavior of stochastic
  schroedinger evolutions}, preprint arXiv:0811.1877 .

\bibitem{Bassi5:08}
A.~Bassi and D.~D\"urr, \emph{On the long time behavior of hilbert space
  diffusion}, Europhys. Lett. \textbf{84}, 10005 (2008).

\bibitem{Ghirardi:86}
G.~C. Ghirardi, A.~Rimini and T.~Weber, \emph{Unified dynamics for microscopic
  and macroscopic systems}, Phys. Rev. D \textbf{34}, 470 (1986).

\bibitem{Adler3:07}
S.~L. Adler, \emph{Lower and upper bounds on {CSL} parameters from latent image
  formation and {IGM} heating}, J. Phys. A \textbf{40}, 2935 (2007).

\bibitem{Ghirardi2:90}
G.~C. Ghirardi, P.~Pearle and A.~Rimini, \emph{Markov processes in hilbert
  space and continuous spontaneous localization of systems of identical
  particles}, Phys. Rev. A \textbf{42}, 78 (1990).

\bibitem{Lipster:01}
R.~S. Liptser and A.~N. Shiryaev, \emph{Statistics of random processes. {I}
  general theory}, volume~5 of \emph{Applications of Mathematics},
  Springer-Verlag, Berlin (2001).

\bibitem{Diosi:97}
L.~Di{\'o}si and W.~T. Strunz, \emph{The non-{M}arkovian stochastic
  {S}chr\"odinger equation for open systems}, Phys. Lett. A \textbf{235}, 569
  (1997).

\bibitem{Diosi:98}
L.~Di{\'o}si, N.~Gisin and W.~T. Strunz, \emph{Non-{M}arkovian quantum state
  diffusion}, Phys. Rev. A \textbf{58}, 1699 (1998).

\bibitem{Bassi:02}
A.~Bassi and G.~C. Ghirardi, \emph{Dynamical reduction models with general
  gaussian noises}, Phys. Rev. A \textbf{65}, 042114 (2002).

\bibitem{Gikhman:96}
I.~I. Gikhman and A.~V. Skorokhod, \emph{Introduction to the theory of random
  processes}, Dover Publications Inc., Mineola, NY (1996).

\bibitem{Yu:99}
T.~Yu, L.~Diosi, N.~Gisin and W.~T. Strunz, \emph{Non-Markovian quantum-state diffusion: perturbation approach}, Phys. Rev. A \textbf{60}, 91 (1999).

\bibitem{Adler:07}
S.~L. Adler and A.~Bassi, \emph{Collapse models with non-white noises}, J. Phys. A \textbf{40}, 15083 (2007).

\bibitem{Adler:08}
S.~L. Adler and A.~Bassi, \emph{Collapse models with non-white noises ii:
  particle-density coupled noises}, J. Phys. A \textbf{41},
  395308 (2008).

\bibitem{Feynman:63}
R.~P. Feynman and Jr.~F.L Vernon, \emph{The theory of a general quantum system interacting with a linear dissipative system}, Ann. Phys. \textbf{24}, 118 (1963).

\bibitem{Pearle:96}
P.~Pearle, in \emph{Perspectives on quantum reality}, Kluwer Acad. Publ., Dordrecht (1996).

\bibitem{Pearle:96bis}
P.~Pearle, in \emph{Open systems and measurement in relativistic quantum theory}, Springer-Verlag, Berlin (1999).

\bibitem{Strunz:96}
W.~T. Strunz, \emph{Linear quantum state diffusion for non-{M}arkovian open
  quantum systems}, Phys. Lett. A \textbf{224}, 25 (1996).

\bibitem{Bernstein:05}
D.~S. Bernstein, \emph{Matrix mathematics}, Princeton University Press,
  Princeton, NJ (2005).

\bibitem{Feynman:65}
R.~P. Feynman and A.~R. Hibbs, \emph{Quantum mechanics and path integrals},
  McGraw-Hill, New York (1965).

\bibitem{Khandekar:83}
D.~C. Khandekar, S.~V. Lawande and K.~V. Bhagwat, \emph{Path integration of a
  two-time quadratic action}, J. Phys. A \textbf{16}, 4209 (1983).

\bibitem{Grosche:98}
C.~Grosche and F.~Steiner, \emph{Handbook of Feynman path integrals}, Springer
  Verlag, Berlin (1998).

\bibitem{Muller:06}
H.~J.~W. M{\"u}ller-Kirsten, \emph{Introduction to quantum mechanics:
  Schr{\"o}dinger equation and path integral}, World Scientific, Hackensack, NJ
  (2006).

\bibitem{Polyanin:03}
A.~D. Polyanin and V.~F. Zaitsev, \emph{Handbook of exact solutions for
  ordinary differential equations}, Chapman \& Hall/CRC, Boca Raton, FL (2003).

\bibitem{Kolokoltsov:98}
V.~N. Kolokol{\cprime}tsov, \emph{Localization and analytic properties of the
  solutions of the simplest quantum filtering equation}, Rev. Math. Phys.
  \textbf{10}, 801 (1998).

\bibitem{Polyanin:08}
A.~D. Polyanin and A.~V. Manzhirov, \emph{Handbook of integral equations},
  Chapman \& Hall/CRC, Boca Raton, FL (2008).

\bibitem{Hartman:02}
P.~Hartman, \emph{Ordinary differential equations}, SIAM, Philadelphia, PA
  (2002).

\bibitem{Ghirardi:90}
G.~C. Ghirardi, R.~Grassi and P.~Pearle, \emph{Relativistic dynamical reduction
  models: general framework and examples}, Found. Phys. \textbf{20}, 1271
  (1990).

\bibitem{Bassi3:05}
A.~Bassi, E.~Ippoliti and B.~Vacchini, \emph{On the energy increase in
  space-collapse models}, J. Phys. A \textbf{38}, 8017 (2005).

\bibitem{Adler2:07}
S.~L. Adler and F.~M. Ramazano{\v{g}}lu, \emph{Photon-emission rate from atomic
  systems in the {CSL} model}, J. Phys. A \textbf{40}, 13395 (2007).

\bibitem{Kreyszig:89}
E.~Kreyszig, \emph{Introductory functional analysis with applications}, Wiley
  Classics Library, John Wiley \& Sons Inc., New York (1989).

\end{thebibliography}
\end{document}